\documentclass[11pt]{article}
\usepackage{mathtools}
\usepackage{amssymb}
\usepackage[mathscr]{euscript}
\usepackage{dsfont}
\usepackage{mathrsfs}
\usepackage{slashed}
\usepackage{fullpage}
\usepackage{currvita}
\usepackage{tocloft}
\usepackage{xcolor}
\usepackage[numbers,sort,compress]{natbib}
\usepackage{array}
\usepackage{enumitem}
\usepackage[normalem]{ulem}
\usepackage{footmisc}
\usepackage[colorlinks=true,linkcolor=blue,citecolor=magenta,linktocpage=true]{hyperref}
\usepackage{titlesec}

\titleformat*{\section}{\normalsize\bfseries}
\titleformat*{\subsection}{\normalsize\bfseries}
\titleformat*{\subsubsection}{\normalsize\bfseries}
\makeatletter
\renewcommand{\@dotsep}{1000}
\def\be#1\ee{\begin{align}#1\end{align}}
\def\bsub#1\esub{\begin{subequations}#1\end{subequations}}
\def\nn{\nonumber}
\def\q{\qquad}
\def\f{\frac}
\def\eps{\varepsilon}

\def\pe{\phantom{\ =}}

\def\de{\mathrm{d}}
\def\A{\mathcal{A}}

\def\D{\mathcal{D}}
\def\E{\mathcal{E}}

\def\I{\mathcal{I}}
\def\J{\mathcal{J}}
\def\K{\mathcal{K}}

\def\M{\mathcal{M}}
\def\N{\mathcal{N}}
\def\O{\mathcal{O}}
\def\P{\mathcal{P}}

\def\T{\mathcal{T}}

\def\W{\mathcal{W}}

\newcommand{\cT}{{\cal T}}

\newcommand{\cW}{{\cal W}}

\renewcommand{\theequation}{\thesection.\arabic{equation}}\numberwithin{equation}{section}
\begin{document}

\title{\Large{\textbf{\sffamily Asymptotically-FLRW$_{\boldsymbol{3}}$ spacetimes}}}
\author{\sffamily Andrea Campoleoni$^1$\footnote{Research associate of the Fonds de la Recherche Scientifique -- FNRS, Belgium.} , Arnaud Delfante$^{1,2}$, Marc Geiller$^3$, Nicolas Maindiaux$^1$\footnote{Research fellow of the Fonds de la Recherche Scientifique -- FNRS, Belgium.}}
\date{}
\date{\small{\textit{
$^1$Service de Physique de l’Univers, Champs et Gravitation,\\ Université de Mons -- UMONS, 20 place du Parc, 7000 Mons, Belgium\\
$^2$Mathematical Physics, Faculty of Physics, University of Vienna,\\ Boltzmanngasse 5, 1090, Vienna, Austria\\
$^3$ENS de Lyon, CNRS, LPENSL, UMR 5672, 69342 Lyon cedex 07, France\\
}}}

\maketitle

\begin{abstract}
We introduce three-dimensional asymptotically-FLRW spacetimes as a simplified setting in which to study asymptotic symmetries and radiation in cosmology. Their asymptotic symmetry group is $\text{BMS}_3^k$, a one-parameter deformation of $\text{BMS}_3$ controlled by the matter equation of state with parameter $k$, in line with the four-dimensional construction of Bonga and Prabhu. We analyze in detail the case of a scalar field matter source, which allows us to fully characterize the solution space and the boundary charges. In particular, we point out that the proper identification of the Bondi mass and angular momentum aspects in the metric requires a careful analysis which had not been laid out so far, even in the existing four-dimensional literature. When superrotations are present, the model exhibits subtleties similar to those appearing when dealing with ``generalized BMS'' asymptotic symmetries in the four-dimensional case, and this requires a covariant definition of the news. We identify covariant notions of news, as well as of mass and angular momentum aspects by studying the vacuum structure, namely the orbits of the vacuum solution under finite $\text{BMS}_3^k$ transformations, and study the Wald--Zoupas prescription for the charges. We also show that these covariant aspects naturally appear in the Cotton scalars, which are the three-dimensional analogues of the Weyl scalars. Finally, we use these quantities to provide a first example of exactly conserved non-linear Newman--Penrose charges in three-dimensional gravity.
\end{abstract}

\newpage

\tableofcontents
\bigskip
\hrule

\newpage

%%%%%%%%%%%%%%%%%%%%%%%%%%%%%%%%%%%%%%%%%%%%%%%%%%%%%%%%%%%%%%%%%%%

\section{Introduction}

Asymptotic symmetries in general relativity were initially discovered in the context of asymptotically-flat spacetimes, as part of the pioneering work of Bondi, van der Burg, Metzner and Sachs on the study of gravitational radiation by isolated sources~\cite{Bondi:1960jsa,Bondi:1962px,Bondi:1962rkt,Sachs:1962zza}. Although these so-called BMS symmetries remained a niche topic in mathematical relativity for a long time, they received a surge of attention in the past decade with the discovery~\cite{Strominger:2013jfa,He:2014laa,Strominger:2014pwa} of their relationship with memory effects~\cite{Braginsky:1986ia,1987Natur.327..123B,Christodoulou:1991cr,Blanchet:1992br,Thorne:1992sdb} and soft theorems~\cite{Low:1954kd,Weinberg:1965nx}, and in turn with the ever more promising prospects for detecting gravitational wave memory~\cite{Lasky:2016knh,Favata:2009ii,Wang:2014zls,NANOGrav:2015xuc,Khera:2020mcz,Mitman:2020pbt,Grant:2022bla}. These studies of BMS symmetries have also brought new insights into flat space holography~\cite{Arcioni:2003xx,Arcioni:2003td,Barnich:2010eb}, and in particular fostered the development of the celestial and Carrollian approaches~\cite{Pasterski:2016qvg,Pasterski:2017kqt,Donnay:2020guq,Pasterski:2021raf,Pasterski:2021rjz,Raclariu:2021zjz,Bagchi:2016bcd,Ciambelli:2018wre,Donnay:2022aba,Donnay:2022wvx,Campoleoni:2023fug,Mason:2023mti,Ruzziconi:2026bix}. Asymptotic symmetries in the presence of a cosmological constant, i.e.\ for asymptotically-(A)dS spacetimes, have also received a lot of attention in both three~\cite{Brown:1986nw,Barnich:2012aw,Grumiller:2016pqb,Alessio:2020ioh,Campoleoni:2022wmf,Ciambelli:2023ott,Ciambelli:2024vhy,Arenas-Henriquez:2024ypo} and four spacetime dimensions~\cite{Henneaux:1985tv,Anninos:2010zf,Anninos:2012qw,Ashtekar:2014zfa,Ashtekar:2015lla,Bieri:2015jwa,Compere:2019bua,Compere:2020lrt,Chrusciel:2020rlz,Compere:2023ktn,Compere:2024ekl,Poole:2025cmv,McNees:2025acf}. BMS symmetries and their consequences have also been studied in the case of boundaries at finite distance in the bulk, such as black hole~\cite{Donnay:2015abr,Akhmedov:2017ftb,Carlip:2017xne,Donnay:2018ckb,Rahman:2019bmk,Grumiller:2019fmp,Agrawal:2025fsv} or cosmological horizons \cite{Bagchi:2012xr,Donnay:2019zif}.

While asymptotically-flat spacetimes are the natural arena to describe astrophysical processes such as black holes or neutron star mergers, another case of interest is that of asymptotically-cosmological spacetimes describing perturbations of cosmological backgrounds. Asymptotically-de~Sitter spacetimes, which describe a phase of accelerated expansion in a vacuum background, have been investigated in several works  \cite{Anninos:2010zf,Anninos:2012qw,Ashtekar:2014zfa,Ashtekar:2015lla,Bieri:2015jwa,Compere:2019bua,Compere:2020lrt,Chrusciel:2020rlz,Compere:2023ktn,Compere:2024ekl,Poole:2025cmv,McNees:2025acf}. However, a physically important case that has received less attention is that of Friedmann--Lemaître--Robertson--Walker (FLRW) backgrounds.\footnote{The question of studying asymptotically-FLRW boundary conditions was already raised in 1962 by J. Weber, as can be seen in the discussion at the end of Bergmann's presentation reported in \cite{Infeld:1964vmz}.} Interestingly, spatially flat FLRW spacetimes undergoing a decelerated expansion possess an asymptotic boundary at future null infinity \cite{Harada:2018ikn,Bonga:2020fhx}, and one can hope to study its properties by leveraging the techniques developed for asymptotically-flat spacetimes. This line of research was initiated by Bonga and Prabhu in~\cite{Bonga:2020fhx} (see also~\cite{Kehagias:2016zry} for a previous study), who showed that spatially flat and decelerating asymptotically-FLRW spacetimes admit asymptotic symmetries akin but not isomorphic to the BMS ones. These symmetries were dubbed ``BMS-like'', and can be thought of as a one-parameter deformation of BMS where supertranslations have a conformal weight depending on a parameter encoding the equation of state of the matter sector. The work~\cite{Bonga:2020fhx} focused mostly on the asymptotic symmetries, and did not provide an in-depth analysis of the solution space, the charges and their fluxes. These questions were later investigated in~\cite{Enriquez-Rojo:2021blc,Enriquez-Rojo:2022onp} (see also~\cite{Enriquez-Rojo:2020miw,Enriquez-Rojo:2022ntu,EnriquezRojo:2022swn,wilson2019thesis}), where an explicit gauge fixing using Bondi coordinates was used to solve the asymptotic Einstein equations and to study proposals and properties of the mass and angular momentum aspects.

One of the subtleties arising in the analysis of asymptotically-FLRW spacetimes is that, when expressed in Bondi coordinates, the line elements feature an overall cosmological scale factor \hbox{$a^2=(u+r)^{2k}$} where $u$ is the retarded time, $r$ the radial coordinate, and $k$ a parameter related to the equation of state of the matter sourcing the stress-energy tensor. Transformations of the scale factor itself under asymptotic symmetries were somehow overlooked in the initial analysis of~\cite{Kehagias:2016zry}, which motivated the corrected treatment presented in~\cite{Bonga:2020fhx} that led to the identification of the BMS-like symmetries. Unfortunately, it turns out that another subtlety due to the presence of the scale factor has crept into the analysis of~\cite{Enriquez-Rojo:2020miw,Enriquez-Rojo:2021blc,Enriquez-Rojo:2022onp,Enriquez-Rojo:2022ntu} as well, and led to an improper identification of the Coulombic Bondi mass and angular momentum aspects. Indeed, given that the asymptotically-FLRW line elements are schematically of the form $\de s^2=a^2\,\de\bar{s}^2$, where $\de\bar{s}^2$ is the same ansatz as for asymptotically-flat spacetimes, the Coulombic aspects have been introduced in the above references at the same order in the radial expansion as they enter in asymptotically-flat metrics. For example, since in four dimensions the mass aspect enters $\de\bar{s}^2$ at order $1/r$ in the asymptotically-flat case, it has been identified in $\de s^2$ at order $a^2/r$ (see, e.g., equation (10) in \cite{Enriquez-Rojo:2022onp}). However, it turns out that this does not correspond to the order at which the radial integration constants, which are identified as the Bondi aspects, appear when solving the asymptotic hypersurface Einstein equations à la Tamburino--Winicour \cite{Tamburino:1966zz}. In fact, for generic values of the parameter $k$, it is necessary to consider an expansion of the metric involving powers of both $r$ and $a$ in order to access the radial integration constants. This therefore goes beyond the structure of the solution space considered in~\cite{Enriquez-Rojo:2020miw,Enriquez-Rojo:2021blc,Enriquez-Rojo:2022onp,Enriquez-Rojo:2022ntu}. In short, the origin of this problem is that asymptotically-FLRW spacetimes are not conformally-asymptotically-flat. The goal of the present work is to illustrate these subtleties, along with other salient features, in the case of three-dimensional asymptotically-FLRW spacetimes.

Indeed, three-dimensional gravity has historically provided an insightful laboratory for classical and quantum aspects of gravity, as it captures many of the same structural and conceptual features as four-dimensional gravity while involving considerably fewer technical complications. In particular, asymptotic symmetries of three-dimensional gravity are notably non-trivial and have been studied at length~\cite{Brown:1986nw,Ashtekar:1996cd,Barnich:2006av,Oblak:2016eij,Barnich:2012aw,Barnich:2012rz,Barnich:2014kra,Barnich:2015uva,Bosma:2023sxn,Cotler:2024cia}. Interestingly, although three-dimensional pure gravity exhibits no propagating degrees of freedom, the coupling to matter enables to describe dynamical spacetimes including cosmological solutions~\cite{Garcia-Diaz:2017cpv}. Since the matter fields sourcing these cosmological solutions can be chosen as radiative, this also enables to study the non-trivial features pertaining to radiative spacetimes. In the present work we demonstrate that this strategy also applies to asymptotically-FLRW spacetimes, which possess three-dimensional realizations exhibiting similar conceptual issues as in the four-dimensional case. However, the structural simplifications due to the lower dimensionality enable us to study the detailed construction of the solution space and the properties of the asymptotic symmetries and charges in a rather manageable setup.

More specifically, our goal is to present a detailed study of spatially flat asymptotically-FLRW$_3$ spacetimes coupled to scalar and Maxwell fields. With this matter content, such spacetimes are decelerating, and the asymptotic symmetries, charges, and fluxes can be studied at the future null boundary~$\mathscr{I}^+$. This study gives results that have direct consequences for four-dimensional asymptotically-FLRW$_4$ spacetimes, namely concerning the identification of the Bondi mass and angular momentum aspects in the solution space. In the three-dimensional case, while the mass aspect enters the metric of asymptotically-flat spacetimes at order $r^0$, we find that it enters the metric of asymptotically-FLRW$_3$ spacetimes at order $a=(u+r)^k$. Therefore, this is not simply given by the asymptotically-flat result dressed by an overall cosmological conformal factor, as one could have initially thought by transposing the ansatz of~\cite{Enriquez-Rojo:2021blc,Enriquez-Rojo:2022onp} to the three-dimensional case. A similar result holds for the angular momentum aspect. This proper identification of the Coulombic data is crucial in order to obtain finite charges and fluxes on $\mathscr{I}^+$. In the case of generic values of the parameter $k$ (in particular if $k$ is non-integer), in order to access the radial integration constants in the metric, the latter should be expanded in powers of both the radius $r$ and the scale factor~$a$. The use of this adjusted ansatz and of its consequences in the four-dimensional case will be the subject of future work.

With the present model, we obtain several additional results which are specific to three spacetime dimensions, but which are nonetheless closely related to results and technical subtleties appearing in the four-dimensional case. 
We show that the asymptotic symmetries of asymptotically-FLRW$_3$ spacetimes are given by a three-dimensional version of the BMS-like symmetries discovered in~\cite{Bonga:2020fhx}. The resulting algebra, denoted by $\mathfrak{bms}_3^k$, is given in \eqref{bmsk algebra} and is a one-parameter deformation of the $\mathfrak{bms}_3$ algebra of asymptotically-flat spacetimes \cite{Ashtekar:1996cd,Barnich:2006av}. We build an explicit realization of this algebra with the asymptotic charges arising from a solution space containing first a radiative scalar field and then the dual Maxwell field. Interestingly, we find that in order to have non-trivial superrotations as part of the asymptotic symmetries, the radial expansion of the scalar field must contain an overleading term with respect to the radiative scalar shear. The presence of this extra mode leads to intricacies in the study of the solution space and of the charges which are exactly analogous to the complications which appear when considering the generalized BMS group in four-dimensional asymptotically-flat spacetimes \cite{Campiglia:2020qvc,Compere:2018ylh}. In particular, this requires to work with an improved and covariant definition of the news $\hat{N}$. In order to arrive at this definition, we proceed by analogy with the four-dimensional work of \cite{Compere:2016jwb,Compere:2018ylh}, and construct the orbit of the gravitational vacua obtained when acting with finite BMS$_3^k$ transformations. By doing so we identify a ``superboost'' field in the solution space, and use it to construct covariant improvements of the mass and angular momentum aspects. These functionals are then used to build Wald--Zoupas charges which are conserved in the radiative vacuum defined by the $\mathfrak{bms}_3^k$-invariant condition $\hat{N}=0$. Although we find that these charges lead to a field-dependent cocycle under the Barnich--Troessaert bracket, there are remaining ambiguities which we have not fixed and which leave room for further improvements, and this clearly indicates that the present model is an ideal testbed to study the Wald--Zoupas and covariance prescriptions recently put forward in \cite{Odak:2022ndm,Rignon-Bret:2024wlu,Rignon-Bret:2024gcx}. We keep this refined investigation for future work.

As an interesting side result and a confirmation of this construction using the superboost field, we also show how the covariant news, mass and angular momentum aspects naturally arise when computing the Cotton scalars. These are indeed the three-dimensional analogues of the four-dimensional Weyl scalars, which also possess at leading order interesting covariance properties \cite{Freidel:2021qpz}. We are then able to recast the evolution equations for the covariant aspects in a form which is analogous to the recursion relation which has been unraveled in four-dimensional asymptotically-flat spacetimes and used to build a celestial $w_{1+\infty}$ algebra \cite{Freidel:2021qpz,Freidel:2021ytz,Geiller:2024bgf}. Finally, this also provides a simple three-dimensional example of exactly conserved Newman--Penrose charges \cite{Newman:1965ik,Newman:1968uj}, which here are built from the modes in the subleading radial expansion of the scalar field. In summary, this three-dimensional model of asymptotically-FLRW spacetimes contains numerous subtleties which are relevant for the understanding of four-dimensional asymptotically-FLRW but also of asymptotically-flat spacetimes.

\newpage

This work is organized as follows. We begin in section \ref{sec:exactly-FLRW} by recalling basic properties of spatially flat exact cosmological solutions to three-dimensional gravity. This enables in particular to introduce Bondi coordinates, to find the dependence of the scale factor and the line element on the deceleration parameter, and to confirm that decelerating spacetimes admit a future null infinity~$\mathscr{I}^+$.

In section \ref{sec:asymptotically-FLRW} we introduce the notion of asymptotically-FLRW$_3$ spacetimes. For this, we give an ansatz in Bondi gauge motivated by the analysis of the exact cosmological solutions. We then exhibit fall-off conditions leading to the $\mathfrak{bms}_3^k$ algebra of residual symmetries, and finally explain how to organize and solve the field equations for a generic matter stress-energy tensor.

Section \ref{sec:scalar} is devoted to the detailed analysis of the coupling to a radiative massless scalar field. This is done in two parts. First, we consider fall-off conditions for the scalar field which only lead to supertranslations and to trivial superrotations (i.e. only spatial rotations). Second, we consider more relaxed fall-off conditions so as to obtain non-trivial superrotations. This is done at the expense of introducing a new field in the solution space, which leads to numerous complications. In particular, the naive news does not transform homogeneously under the action of the asymptotic symmetries, which prevents from defining a radiative vacuum and Wald--Zoupas charges. In order to address this, we study the vacuum structure and identify a covariant news. We also introduce covariant mass and angular momentum aspects, and show that they appear in the expansion of the Cotton scalars. We then compute the charge algebra and show that it leads to $\mathfrak{bms}_3^k$ with a field-dependent cocycle. We end this section by analyzing the subleading structure of the solution space, which shows that some of the evolution equations in retarded time can be rewritten in terms of the Cotton scalars. We also identify an exactly conserved Newman--Penrose charge built from the subleading expansion of the scalar field.

In section \ref{sec:Maxwell} we repeat part of the analysis of the scalar case for a Maxwell field. This is done by exploiting the duality which exists between a scalar and a Maxwell field in three-dimensional spacetimes. We finally conclude and give perspectives for future work in section \ref{sec:conclusion}. The appendices contain reminders about three-dimensional asymptotically-flat spacetimes with and without matter, remarks about the extension of the $\mathfrak{bms}_3^k$ algebra to include Weyl transformations, about the absence of a translation subalgebra, lengthy equations and expressions for the bare charges, and finally details about the computation of the vacuum structure.

%%%%%%%%%%%%%%%%%%%%%%%%%%%%%%%%%%%%%%%%%%%%%%%%%%%%%%%%%%%%%%%%%%%

\section{Exactly-FLRW\texorpdfstring{$_{\boldsymbol{3}}$}{3} spacetimes}
\label{sec:exactly-FLRW}

To begin with, it is useful to recall some basic results on the structure of cosmological models in three spacetime dimensions. While three-dimensional vacuum gravity is topological and admits no propagating degrees of freedom, coupling to matter gives rise to dynamical spacetimes, including FLRW solutions (see, e.g., \cite{Garcia-Diaz:2017cpv} for a review). In this section, we briefly review the form of spatially flat cosmological solutions in various coordinate systems, and in particular introduce the Bondi coordinates that will be used in the analysis of the asymptotic structure.

\newpage

We consider a perfect fluid with equation of state $p=w\rho$, with $w$ a constant relating the pressure $p$ to the energy density $\rho$. Its dynamics is encoded in the stress-energy tensor
\be \label{EMT}
T_{\mu\nu}=\big(p+\rho\big)U_\mu U_\nu+pg_{\mu\nu},
\q\q
U^\mu U_\mu=-1,
\ee
where $g_{\mu\nu}$ is the spacetime metric and $U^\mu$ is a normalized velocity vector. The metric and the stress-energy tensor satisfy the field equations $E_{\mu\nu}\coloneqq G_{\mu\nu}-\kappa T_{\mu\nu}=0$, where $G_{\mu\nu}$ is the Einstein tensor, together with the conservation law $\nabla^\mu T_{\mu\nu}=0$. In what follows we set $\kappa=8\pi G=1$ for conciseness.

\paragraph{Polar coordinates.}

We begin by introducing polar coordinates $x^\mu = (t,r,\phi)$ and the spatially flat, homogeneous, and isotropic FLRW ansatz
\be\label{exact FLRW polar}
\de s^2=-\de t^2+a(t)^2\big(\de r^2+r^2\de\phi^2\big),
\q\q
U_\mu=-(1,0,0),
\ee
where $a(t)$ is the scale factor. With this metric the conservation law reduces to
\be\label{exact FLRW cons energy}
\nabla^\mu T_{\mu t}=0\quad\Rightarrow\quad\f{\dot{\rho}}{\rho}=-2(1+w)\f{\dot{a}}{a},
\ee
and the independent components of the Einstein equations imply
\be\label{Exact FLRW Einstein}
E_{tt}=0\quad\Rightarrow\quad\left(\f{\dot{a}}{a}\right)^2=\rho,
\q\q\q
E_{rr}=r^2E_{\phi\phi}=0\quad\Rightarrow\quad\f{\ddot{a}}{a}=-w\rho,
\ee
where the dot denotes $\partial_t$. For $w\neq-1$, solving \eqref{exact FLRW cons energy} leads to
\be\label{rho FLRW}
\rho=\mathrm{c}^{2(1+w)}\f{a^{-2(1+w)}}{(1+w)^2},
\ee
where $\mathrm{c}$ is an integration constant which we set to $\mathrm{c}=1$ for later convenience and without loss of generality. From \eqref{Exact FLRW Einstein} we then get that the scale factor exhibits, as expected, a monotonic dependence on the Cartesian time $t$. For the expanding branch (the collapsing branch can be obtained by time reversal), solving \eqref{Exact FLRW Einstein} with a suitable choice for the additional integration constant leads~to
\be\label{cosmic a}
\begin{cases}
a(t)=(+t)^{1/(1+w)}\q\quad\text{for $1+w>0$ and $0<t<+\infty$},\\
a(t)=(-t)^{1/(1+w)}\q\quad\text{for $1+w<0$ and $-\infty<t<0$}.
\end{cases}
\ee
The limiting case $w=-1$ between these two regimes corresponds to dS$_3$, where $\rho$ is constant and $a\sim e^t$. AdS$_3$ cannot instead be described with the initial ansatz \eqref{exact FLRW polar} for the line element, while Minkowski spacetime can be obtained from the limit $w\to\infty$. Finally, we note that in these coordinates the deceleration parameter is
\be
q=-\f{a\ddot{a}}{\dot{a}^2}=w.
\ee
The solution is said to be decelerating when $w>0$, and accelerating when $w<0$.

It is interesting to also briefly discuss the isometries of these solutions. The spatial homogeneity and isotropy of the three-dimensional FLRW line element \eqref{exact FLRW polar} imply the existence of exact Killing vectors generating spatial translations and rotations. In polar coordinates, these are generated respectively by
\be\label{exact KV polar}
\xi_{T_1}=\sin\phi\,\partial_r+\f{\cos\phi}{r}\,\partial_\phi,
\q\q
\xi_{T_2}=\cos\phi\,\partial_r-\f{\sin\phi}{r}\,\partial_\phi,
\q\q
\xi_{Y_0}=\partial_\phi.
\ee
They span the three-dimensional algebra $\mathfrak{iso}(2)$, whose non-vanishing brackets are $[\xi_{Y_0},\xi_{T_i}]=\epsilon_{ij}\xi_{T_j}$ with $i\in\{1,2\}$ and $\epsilon_{12}=1$. For $w\neq-1$, these are all Killing vectors of the metric \eqref{exact FLRW polar}. More generally, for a $d$-dimensional FLRW metric the Killing vectors correspond only to the symmetries of the maximally-symmetric $(d-1)$-dimensional spatial slices, and do not contain time translations. This remark is important since we will nonetheless obtain supertranslations generators for our asymptotically-FLRW spacetimes, in analogy with what has been achieved in four dimensions \cite{Kehagias:2016zry,Bonga:2020fhx,Enriquez-Rojo:2020miw}. Contrary to what happens for asymptotically-flat spacetimes, where supertranslations are an infinite-dimensional enhancement of the translations of the background Minkowski spacetime, for asymptotically-FLRW spacetimes one cannot strictly speaking refer to an enhancement since time translations are absent in the symmetries of the background. We will come back to this point below when discussing the asymptotic symmetries.

\paragraph{Conformal coordinates.}

We now switch to conformal coordinates $x^\mu=(\tau,r,\phi)$, where the conformal time is defined from the cosmic time as $\de\tau=\de t/a(t)$. The line element and the velocity take the form
\be\label{conformal coords}
\de s^2=a(\tau)^2\big(\!-\!\de\tau^2+\de r^2+r^2\de\phi^2\big),
\q\q
U_\mu=-a(\tau)(1,0,0).
\ee
For $w\neq-1$, readjusting conveniently the integration constants, the conservation and Einstein equations are solved in the expanding branch by
\be\label{conf a}
\rho=\f{a^{-2(1+w)}}{w^2},
\q\q
\begin{cases}
a(\tau)=(+\tau)^{1/w}\q\quad\text{for $w>0$ and $0<\tau<+\infty$},\\
a(\tau)=(-\tau)^{1/w}\q\quad\text{for $w<0$ and $-\infty<\tau<0$}.
\end{cases}
\ee
With these coordinates, the deceleration parameter is\footnote{In four-dimensional FLRW spacetimes filled with a perfect fluid we have instead that $q=(1+3w)/2$, so that deceleration corresponds to $w>-1/3$.}
\be
q=1-\f{a\ddot{a}}{\dot{a}^2}=w,
\ee
where the dot now denotes $\partial_\tau$.

\paragraph{Bondi coordinates.}

Finally, let us define the retarded time coordinate $u=\tau-r$, which puts the line element and the velocity in the conformal Bondi form
\be\label{Pure FLRW bondi coord.}
\de s^2=a(u,r)^2\big(\!-\!\de u^2-2\de u\,\de r+r^2\de\phi^2\big),
\q\q
U_\mu=-a(u,r)(1,1,0).
\ee
Keeping the same normalizations as in \eqref{conf a}, the solutions to the conservation and Einstein equations are
\be\label{Bondi rho and a}
\rho=\f{a^{-2(1+w)}}{w^2},
\q\q
a=(u+r)^{1/w},
\ee
while the deceleration parameter remains $q=w$. In what follows we will denote
\be\label{definition k}
k\coloneqq\f{1}{w},
\ee
so that the scale factor reads $a=(u+r)^k$. The cosmological line element \eqref{Pure FLRW bondi coord.} is given by the Minkowski metric in retarded Bondi coordinates multiplied by the dynamical scale factor. Below we shall use a similar ansatz to construct asymptotically-FLRW$_3$ metrics which, as their four-dimensional counterparts \cite{Kehagias:2016zry, Bonga:2020fhx, Enriquez-Rojo:2020miw, Enriquez-Rojo:2021blc, Enriquez-Rojo:2022onp, Enriquez-Rojo:2022ntu}, will be given by asymptotically-flat metrics in Bondi gauge multiplied by the scale factor. In Bondi coordinates, the exact Killing vectors \eqref{exact KV polar} of the FLRW$_3$ background become
\be\label{exact Killing FLRW Bondi}
\xi_{T_1}=\sin\phi\,\partial_u-\sin\phi\,\partial_r-\f{\cos\phi}{r}\, \partial_\phi,
\q\;
\xi_{T_2}=\cos\phi\,\partial_u-\cos\phi\,\partial_r+\f{\sin\phi}{r}\, \partial_\phi,
\q\;
\xi_{Y_0}=\partial_\phi.
\ee
Note that the specific form of $a(u,r)$ in \eqref{Bondi rho and a} is required to show that $\xi_{T_i}$ are Killing vectors.

It is also instructive to study the fall-offs of the stress-energy tensor when $r\to\infty$ in Bondi coordinates. Inserting \eqref{Pure FLRW bondi coord.} and \eqref{Bondi rho and a} in \eqref{EMT} and using $w=1/k$, we find that the components are given by
\bsub\label{background Tmunu}
\be
T_{uu}&=T_{ur}=\f{T_{rr}}{1+k^{-1}}=\f{k^2}{(u+r)^2}=\f{k^2}{r^2}+\O(r^{-3}),\\
T_{\phi\phi}&=\f{kr^2}{(u+r)^2}=k\left(1-\f{2u}{r}+\f{3u^2}{r^2}\right)+\O(r^{-3}),\\[5pt]
T_{u\phi}&=T_{r\phi}=0.
\ee
\esub
This provides an indication of which minimal fall-offs for the stress-energy tensor should be considered when constructing the general solution space.

\paragraph{Conformal diagrams.}

Lastly, let us discuss the conformal diagrams of these solutions in order to explain why we will focus on decelerating FLRW$_3$ spacetimes, i.e.\ on the case $w > 0$. Indeed, this is the only case in which the spacetimes admit a region identified with future null infinity, and which is reached when sending $r\to\infty$ while keeping $u$ constant. The proof is analogous to the four-dimensional counterpart presented in \cite{Kehagias:2016zry,Bonga:2020fhx}. In Bondi coordinates, the retarded time $u$ is related to the conformal time $\tau$ via $u=\tau-r$. In order to send $r\to\infty$ while keeping $u$ constant we must therefore send~$\tau\to\infty$ as well. From \eqref{conf a}, one can see that this is only possible when $w>0$, i.e.\ for decelerating spacetimes. The same conclusion holds in the contracting branch.

\begin{figure}[ht!]
\centering
\includegraphics[width=7cm]{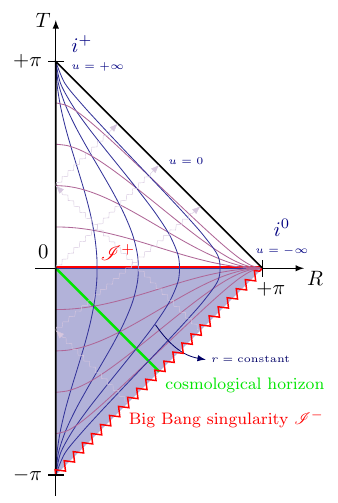}\q\includegraphics[width=7cm]{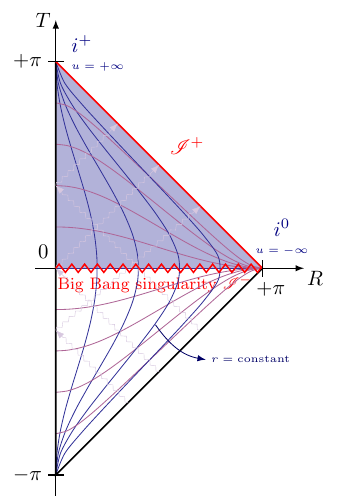}
\caption{Penrose diagrams of spatially flat accelerating (left) and decelerating (right) FLRW spacetimes, in shaded purple. Adapted from \cite{neutelings2024}.}
\label{Pen diagr dec}
\end{figure}

The conformal diagrams of the accelerating and decelerating spacetimes are represented in figure~\ref{Pen diagr dec}. On these diagrams, the spacetimes have been compactified by using coordinates $(T,R)$ defined from the conformal coordinates \eqref{conformal coords} and \eqref{conf a} as
\be
\tau=\f{\sin T}{\cos R+\cos T},
\q\q
r=\f{\sin R}{\cos R+\cos T}.
\ee
In the expanding branch, these coordinates have a finite range given by
\be
\text{$-\pi<T<0$ and $0<R<\pi+T$ when $-\infty<\tau<0$ and $r>0$, i.e. $w<0$},\cr
\text{$0<T<+\pi$ and $0<R<\pi-T$ when $0<\tau<+\infty$ and $r>0$, i.e. $w>0$},\nn
\ee
and they put the line element in the conformally-compactified form
\be
\de s^2=\Omega^2\big(\!-\!\de T^2+\de R^2+\sin^2R\,\de\phi^2\big),
\q\q
\Omega\coloneqq\f{(\sin|T|)^{1/w}}{2\left(\cos\f{R+T}{2}\,\cos\f{R-T}{2}\right)^{1+1/w}}.
\ee
A crucial feature of this conformal compactification is that, depending on the sign of $w$, the finite range of the coordinates implies that only a portion of the conformal diagram of empty Minkowski spacetime is recovered as the physical portion of the asymptotically-FLRW spacetimes. Only the decelerated case ($w>0$) admits a future null infinity $\mathscr{I}^+$, while in the accelerated case ($w<0$) $\mathscr{I}^+$ is spacelike and there is a null cosmological horizon. For a scalar or Maxwell field we have $w=1$, and it is therefore possible to study future null infinity by adapting the techniques developed for asymptotically-flat spacetimes.

%%%%%%%%%%%%%%%%%%%%%%%%%%%%%%%%%%%%%%%%%%%%%%%%%%%%%%%%%%%%%%%%%%%

\section{Asymptotically-FLRW\texorpdfstring{$_{\boldsymbol{3}}$}{3} spacetimes}
\label{sec:asymptotically-FLRW}

We now turn to the study of asymptotically-FLRW$_3$ spacetimes. To characterize them, we first specify a coordinate gauge and then impose boundary conditions on the fluctuating components of the metric. More precisely, we consider radial fall-off conditions that allow for residual diffeomorphisms mimicking asymptotically-flat supertranslations and superrotations~\cite{Barnich:2006av}. This construction leads to a three-dimensional analogue of the four-dimensional results of~\cite{Bonga:2020fhx}, namely a $k$-deformed~$\mathfrak{bms}_3$ asymptotic symmetry algebra, denoted $\mathfrak{bms}_3^{k}$, where $k$ is the factor defined in~\eqref{definition k} and characterizing the matter equation of state.

After discussing the residual symmetries, we also present in this section an algorithm for solving the field equations $E_{\mu\nu}\coloneqq G_{\mu\nu}-T_{\mu\nu} = 0$, following an analogue of the Bondi hierarchy à la Tamburino--Winicour \cite{Tamburino:1966zz} adapted to the presence of a cosmological expansion and a non-vanishing stress-energy tensor. As in standard studies of cosmological perturbations, we stress that, even for fixed $k$, our allowed perturbations over the background solutions described in section~\ref{sec:exactly-FLRW} are such that, in general, the matter sector is not described anymore by a perfect fluid. In fact, one could even contemplate studying the case of a time-dependent $k$, but this technically challenging task is beyond the scope of the present paper.

We stress that from the purely algebraic viewpoint the results presented in this section hold for any value of $k$, i.e. for decelerating as well as accelerating cosmological spacetimes. However, as explained above and in figure \ref{Pen diagr dec}, the global geometry depends on $k$ and only the decelerated case admits a future null infinity.

\subsection{Gauge conditions, fall-offs, and residual diffeomorphisms}
\label{sec:gauge and fall-offs}

Guided by the form~\eqref{Pure FLRW bondi coord.} of the FLRW$_3$ line element in Bondi coordinates, and by the characterization of asymptotically-flat spacetimes developed in~\cite{Barnich:2010eb}, we work in a gauge in which the metric takes the form
\be\label{metric ansatz}
\de s^2=a^2\left(\f{V}{r}e^{2\beta}\de u^2-2e^{2\beta}\de u\,\de r+r^2(\de\phi-U\de u)^2\right),
\q
a=(u+r)^k,
\ee
where $(\beta,U,V)$ are functions of all coordinates $(u,r,\phi)$. This line element satisfies the standard Bondi gauge conditions $g_{rr}=0=g_{r\phi}$, as well as $g_{\phi\phi}=a^2r^2$ which generalizes the determinant condition. The background solution \eqref{Pure FLRW bondi coord.} is recovered for $V=-r$ and $\beta=0=U$. We now have to specify the leading order behavior of the functions $(\beta,U,V)$ when $r\to\infty$. Our goal is to identify fall-offs admitting residual diffeomorphisms generated by two arbitrary functions on the celestial circle, which will be the supertranslations and superrotations of the $\mathfrak{bms}_3^k$ algebra. At the same time, we want these fall-offs to be compatible with the asymptotic form of the metric arising from the resolution of the Einstein equations $E_{\mu\nu}=0$ sourced by a non-trivial stress-energy tensor. It turns out that such a choice of boundary conditions is given by
\be\label{fall-offs beta V u}
\beta=\O(r^{-1}),
\q\q
V=\O(r),
\q\q
U=\O(r^{-1}),
\ee
or equivalently for the metric components
\be\label{fall-offs metric}
a^{-2}g_{uu}=\O(1),
\q\q
a^{-2}g_{ur}=-1+\O(r^{-1}),
\q\q
a^{-2}g_{u\phi}=\O(r).
\ee
Besides the overall factor of $a(u,r)$ in the metric, these fall-offs can be contrasted with the boundary conditions \eqref{flat vacuum BCs} used for three-dimensional vacuum asymptotically-flat spacetimes, where one has in particular the stronger radial constraint $U=\O(r^{-2})$. In our context, imposing the latter condition would kill the superrotations.

We can now characterize the residual diffeomorphisms which preserve the form \eqref{metric ansatz} of the line element as well as the fall-off conditions \eqref{fall-offs metric}. First, in order to preserve the gauge choice, we search for vector fields such that
\be\label{preserving the gauge}
\pounds_\xi g_{rr}=0,
\q\q
\pounds_\xi g_{r\phi}=0,
\q\q
\pounds_\xi g_{\phi\phi}=0,
\ee
where $\pounds$ denotes the Lie derivative. One should note that the last condition derives from the fact that the scale factor in $g_{\phi\phi}=a^2r^2$ depends only on the coordinates and not on the metric fluctuations. Solving these constraints leads to vector fields of the form
\bsub\label{AKV components}
\be
\xi^u&=f(u,\phi),\\
\xi^\phi&=y(u,\phi)-f'\int_r^\infty\f{e^{2\beta}}{\tilde{r}^2}\de\tilde{r},\\
\xi^r&=\f{r}{1+r\partial_r\ln a}\Big(Uf'-f\partial_r\ln a-(\xi^\phi)'\Big),
\ee
\esub
where a prime denotes $\partial_\phi$ and where $\partial_r\ln a=\partial_u\ln a=k(r+u)^{-1}$. At this stage, $f(u,\phi)$ and~$y(u,\phi)$ are two arbitrary functions of the retarded time and the angular coordinate. On top of the gauge conditions, one has to impose that the diffeomorphism preserves the fall-offs \eqref{fall-offs metric}. First, we find that preserving the third boundary condition leads to
\be\label{preserving g u phi}
\pounds_\xi g_{u\phi}=a^2\O(r)\q\Rightarrow\q y(u,\phi)=Y(\phi).
\ee
With this constraint we then find that the first constraint in \eqref{fall-offs metric} is automatically preserved, i.e.~$\pounds_\xi g_{uu}=a^2\O(1)$. Finally, preserving the second boundary condition leads to
\be
\pounds_\xi g_{ur}=a^2\O(r^{-1})\q\Rightarrow\q f(u,\phi)=T(\phi)+u\f{1+2k}{1+k}Y'(\phi).
\ee
At the end of the day we are therefore left, as announced, with two free functions $T(\phi)$ and $Y(\phi)$ of the angular coordinates, which parametrize respectively supertranslations and superrotations. Consistently, in the limit $k\to0$ (which implies $a\to1$) the above results reproduce the $\mathfrak{bms}_3$ asymptotic Killing vectors of three-dimensional asymptotically-flat spacetimes.\footnote{Strictly speaking, this result requires to study how the metric functions $\beta$ and $U$ entering the vector field \eqref{AKV components} reduce to the standard expressions for asymptotically-flat spacetimes in the case of a vanishing stress-energy tensor.}

Now that we have characterized the vector fields generating the residual diffeomorphisms, we can compute their algebra. Since the vector fields are field-dependent, the study of their algebra requires to use the modified bracket introduced in \cite{Schwimmer:2008yh,Barnich:2010eb}, which in turn requires to know the transformation laws of the various independent data in the solution space. In order to simplify this analysis, one may instead focus on the vector field intrinsic to the asymptotic boundary. This vector field is field-independent and takes the form
\be\label{residual diffeo on scri}
\xi_{(T,Y)}=\left(T+u\f{1+2k}{1+k}Y'\right)\partial_u +Y\partial_\phi.
\ee
Evaluating the Lie bracket, one then finds
\be\label{Lie bracket}
\big[\xi_{(T_1,Y_1)},\xi_{(T_2,Y_2)}\big]=\xi_{(T_{12},Y_{12})},
\ee
with
\be\label{eq. bms3like}
T_{12}\coloneqq Y_1T_2'+\f{1+2k}{1+k}T_1Y_2'-(1\leftrightarrow 2),
\q\q
Y_{12}\coloneqq Y_1Y_2'-(1\leftrightarrow 2).
\ee
This defines the $\mathfrak{bms}_3$-like algebra
\be\label{bmsk algebra}
\mathfrak{bms}_3^k=\mathrm{Vect}(S^1)\ltimes T_k(S^1),
\ee
where the supertranslations $T_k(S^1)$ are functions on the celestial circle $S^1$ with conformal weight $(1 + 2k)/(1 + k)$. This cosmologically-modified $\mathfrak{bms}_3$ algebra is the three-dimensional analogue of the four-dimensional result $\mathfrak{bms}_4^k=\text{Vect}(S^2)\ltimes T_k(S^2)$ obtained in \cite{Bonga:2020fhx,Enriquez-Rojo:2020miw}, the only difference being in the definition of $k_\text{4d}\coloneqq2/(3w+1)\neq k_\text{3d}$. In appendix \ref{app:Weyl} we extend this analysis by enlarging the residual symmetries to include Weyl transformations. Interestingly, this leads to a residual symmetry algebra which is isomorphic to that of asymptotically-flat vacuum gravity.

Expanding \eqref{residual diffeo on scri} in Fourier modes allows one to compare with the vector field induced on the boundary by the exact Killing vectors \eqref{exact Killing FLRW Bondi} of the background. Rotations correspond to a constant $Y(\phi)=Y_0$, while spatial translations correspond to the first two Fourier modes of $T(\phi)$. However, one should note that the constant Fourier mode in the expansion of $T(\phi)$ does not correspond to an exact Killing vector. Considering metric fluctuations therefore brings the option to include asymptotic time translations among the symmetries, even if they are not symmetries of the non-stationary background. In appendix \ref{app:subalgebra} we prove that the full algebra $\mathfrak{bms}_3^k$ in \eqref{bmsk algebra} does not admit a preferred translation subalgebra. However, when restricting the superrotations to only the constant rotations $Y(\phi)=Y_0$, which reduces \eqref{bmsk algebra} to $\mathfrak{so}(2)\ltimes T_k(S^1)$, there does actually exist a subalgebra of translations, and there is even a smaller subalgebra of time translations.

\subsection{Organization of the field equations}
\label{sec:Bondi hierarchy}

We now present the algorithm we use to solve the Einstein equations for the line element~\eqref{metric ansatz}, allowing for a stress-energy tensor that includes fluctuations about its background value. The equations are organized as in the standard Bondi hierarchy \cite{Tamburino:1966zz}, i.e.\ into hypersurface equations which determine the radial behavior of the metric functions $(U,V,\beta)$ entering \eqref{metric ansatz}, evolution equations in retarded time $u$, and trivial equations. Some lengthy explicit expressions for certain components of the Ricci tensor and the Einstein equations are collected in appendix~\ref{app: Einstein eqs}.

The resolution algorithm starts with the first hypersurface equation $E_{rr}=0$, which enables to determine the radial behavior of the function $\beta$ once an expansion for $T_{rr}$ has been given.\footnote{We shall discuss this point in more detail in the following sections, starting with the example of a massless scalar field, corresponding to $k=1$.} The equation is
\be \label{eq. Err}
E_{rr}=R_{rr}-T_{rr}=\f{2(\partial_ra)^2}{a^2}-\f{\partial_r^2a}{a}+2\left(\f{\partial_ra}{a}+\f{1}{r}\right)\partial_r\beta-T_{rr}=0.
\ee
Since it is of first order in $\partial_r\beta$, it allows for a radial integration ``constant'' $\beta_0(u,\phi)$ at order $\O(1)$ in the radial expansion of the function $\beta$, which we however set to zero in order to satisfy the fall-offs on $g_{ur}$ in \eqref{fall-offs metric}.

The second hypersurface equation which must be studied is $E_{r\phi}=0$, which determines the radial behavior for $U$. Explicitly, we have
\be \label{eq. Erphi}
E_{r\phi}=R_{r\phi}-T_{r\phi}=\f{1}{2ar}\partial_r\big(ar^3e^{-2\beta}\partial_rU\big)-ar\partial_r\left(\f{\beta'}{ar}\right)-T_{r\phi}=0.
\ee
This second-order equation in $\partial_r^2 U$ gives rise to two integration constants. The first one also appears at order $\mathcal{O}(1)$ in the radial expansion and we set it to zero in order to satisfy the fall-off conditions~\eqref{fall-offs metric}. By contrast, the second arises at order $1/(a r^{2}) \sim r^{-2-k}$. Since we are considering only decelerating FLRW backgrounds with $k > 0$, it is always compatible with the boundary conditions~\eqref{fall-offs beta V u} and will later be used as the starting point to construct the Coulombic data representing the FLRW analogue of the Bondi angular momentum aspect. We note that this identification contrasts with previous work on four-dimensional asymptotically-FLRW solutions~\cite{Enriquez-Rojo:2022onp}, where the angular momentum aspect was assumed to appear in $U$ at the same radial order as in the vacuum asymptotically-flat case (i.e., at order $1/r^3$ in the expansion of $U$ in the four-dimensional case, which would correspond to an order $1/r^2$ in the three-dimensional case).\footnote{A similar shift appears when studying the linearized equations of motion for massless fields of any spin on a spatially flat three-dimensional FLRW background. In this case (see, e.g., \cite{Campoleoni:2012hp}), one can construct a consistent deformation of the Fronsdal equations describing massless fields of integer spin $s$. Relative to the flat-space analyses of \cite{Gonzalez:2013oaa, Afshar:2013vka}, the order at which Coulombic data appear is however shifted by a power $a$.} We will come back to this subtle point in detail when studying explicit examples below. In short, the conclusion is that the inclusion of angular momentum and mass aspects in the solution space requires to expand the functions entering the metric ansatz \eqref{metric ansatz} not only in integer powers of $r$, but also in powers of the conformal factor $a$.

The remaining hypersurface equation determines the radial expansion of the metric function~$V$. It turns out that the Ricci component $R_{ur}$ entering the $(ur)$ Einstein equation is of second order in~$\partial_r^2V$. For this reason, it is convenient to rewrite $E_{ur}$ taking into account the previous hypersurface equations, $E_{rr}=0=E_{r\phi}$. This can be done by first rewriting the Ricci scalar as
\be
R=2g^{ur}R_{ur}+g^{\phi\phi}R_{\phi\phi}+g^{rr}T_{rr}+2g^{r\phi}T_{r\phi}.
\ee
Together with the fact that $g^{ur}g_{ur}=1$, this leads to a rewriting of $E_{ur}=0$ in the form
\be\label{eq. Eur}
a^2e^{2\beta}g^{\phi\phi}R_{\phi\phi}-\f{V}{r}T_{rr}-2UT_{r\phi}-2T_{ur}=0,
\ee
where only the first-order derivative $\partial_rV$ now enters $R_{\phi\phi}$ given in \eqref{Rphiphi}. Once again, the resolution leads to an integration constant, now at order $r/a$ since, schematically,
\be
R_{\phi\phi} = e^{-2\beta}\f{r}{a} \partial_r\left(\partial_r(ar)\f{V}{r}\right)+(\,\dots)
\ee
which will be used later on to build the mass aspect. This illustrates once again the need to use both a radial and an $a$-dependent expansion for the functions entering the metric. In addition, one should note that solving $E_{ur}=0$ also leads to new constraints involving some of the subleading components of the stress-energy tensor.

We now have to discuss the evolution equations, which constrain in particular the retarded time evolution of the Coulombic data, i.e., of the mass and the angular momentum aspects. At order $1/(ar)$, the equations $E_{u\phi}=0$ and $E_{uu}=0$ given in \eqref{Euphi} and \eqref{Euu} contain respectively the evolution equations for the angular momentum aspect and for the mass aspect. In the absence of matter, one can use the contracted Bianchi identities to show that there are no more equations to be solved, and in particular that $E_{\phi\phi}=0$ is automatically satisfied. Here, however, the presence of a non-vanishing stress-energy tensor prevents from using directly this argument, and there are further equations to be studied and solved. On the other hand, the remaining equations are equivalent to the conservation of the stress-energy tensor, as one can see from the contracted Bianchi identities~(CBI),
\be\label{Bianchi contracted identities}
(\text{CBI})_\nu\coloneqq\partial_\mu\big(\sqrt{-g}\,{E^\mu}_\nu\big)+\f{\sqrt{-g}}{2}E_{\alpha\beta}\partial_\nu g^{\alpha\beta}-\sqrt{-g}\,\nabla_\mu{T^\mu}_\nu\equiv0.
\ee
When the hypersurface equations $E_{r\mu}=0$ are satisfied, $(\text{CBI})_r$ can be rewritten as
\be\label{CBIr}
(\text{CBI})_r\quad\Rightarrow\quad E_{\phi\phi}\partial_rg^{\phi\phi}-2\nabla_\mu{T^\mu}_r=0,
\ee
where $E_{\phi\phi}$ is given in \eqref{Ephiphi}. In the absence of matter, this is the equation which implies that $E_{\phi\phi}$ is trivially satisfied once the hypersurface equations are imposed. Here, by contrast, this equation involves the divergence of the stress-energy tensor or, equivalently, the matter equations of motion. The explicit form of these equations should be worked out case-by-case depending on the matter sector being considered, and we shall give explicit examples below.

We conclude with the remaining two components of the Bianchi identities. On shell, using the equations solved above, they reduce to
\bsub\label{CBIphi and u}
\be
(\text{CBI})_\phi\quad&\Rightarrow\quad\partial_r(arE_{u\phi})+ra^3e^{2\beta}\nabla_\mu{T^\mu}_\phi=0,\\
(\text{CBI})_u\quad&\Rightarrow\quad\partial_r(arE_{uu})+ra^3e^{2\beta} \nabla_\mu T^\mu{}_u=0,
\ee
\esub
where $E_{u\phi}$ is given in~\eqref{Euphi} and $E_{uu}$ in~\eqref{Euu}. In the asymptotically flat vacuum case (where we recall that $a=1$), these two equations show that $E_{u\phi}$ and $E_{uu}$ each contain a single independent term at order~$r^{-1}$, corresponding to the evolution equations for the angular momentum and mass aspects. Here, once the matter equation of motion is satisfied, the situation is similar with the only difference that the non-trivial information appears at order $(ar)^{-1}$.

This completes the analysis of the Einstein equations and the construction of the solution space. Given the fall-off conditions for the metric and the stress-energy tensor, this algorithm determines the radial expansion of the metric functions in terms of the components of the stress-energy tensor, up to the Coulombic data representing the mass and the angular momentum, which emerge as integration ``constants'' for the hypersurface equations \eqref{eq. Erphi} and \eqref{eq. Eur}. It also yields constraints among the components of the stress-energy tensor arising from its conservation, as well as evolution equations for the Bondi-like mass and angular momentum aspects encoded in $E_{uu} = 0$ and $E_{u\phi} = 0$.

%%%%%%%%%%%%%%%%%%%%%%%%%%%%%%%%%%%%%%%%%%%%%%%%%%%%%%%%%%%%%%%%%%%

\section{Coupling to a scalar field}
\label{sec:scalar}

In this section, we undertake a more detailed construction and analysis of asymptotically-FLRW$_3$ spacetimes, focusing on a matter sector described by a massless scalar field whose general features are discussed in section~\ref{subsec: scalar_generalities}. We begin in \ref{subsec: scalar_stranslations} by imposing boundary conditions on the scalar field that lead only to supertranslations and exclude superrotations. This restricted setting allows us to build intuition on how to identify mass and angular momentum aspects, in the spirit of the Bondi-like hierarchy discussed in subsection~\ref{sec:Bondi hierarchy}, without having to deal with the subtleties brought by superrotations. In subsection~\ref{sec:scalar_superrotations}, we then extend the discussion to scalar fall-off conditions that lead to non-trivial superrotations. This extension requires in particular to discuss the vacuum structure and the identification of a proper notion of radiative news defined with a three-dimensional analogue of the Geroch tensor.

\subsection{Generalities} \label{subsec: scalar_generalities}

We consider a massless scalar field with a dynamics arising from $L=L_{\mathrm{EH}}+L_{\mathrm{scalar}}$, where the Einstein--Hilbert and scalar field Lagrangian densities are given by
\be\label{eq. LEH-scalar}
L_\text{EH}=\f{1}{2}\sqrt{-g}\,R,
\q\q
L_\text{scalar}=-\f{1}{4}\sqrt{-g}\,\nabla^\mu\Phi\nabla_\mu\Phi.
\ee
The variation of the total Lagrangian with respect to the metric and the scalar field takes the form
\be \label{eq. deltaL}
\delta L=\sqrt{-g}\,\big(\delta g^{\mu\nu}E_{\mu\nu}+\delta\Phi\,\Box\Phi\big)+\partial_\mu\theta^\mu,
\ee
where the equations of motion are
\be\label{eq. EHscalar eoms}
E_{\mu\nu}\coloneqq G_{\mu\nu}-T_{\mu\nu}=0,
\q\q
\Box\Phi=\nabla^\mu\nabla_\mu\Phi=0.
\ee
The matter stress-energy tensor is given by
\be\label{scalar stress tensor}
T_{\mu\nu}=\f{1}{2}\nabla_\mu\Phi\nabla_\nu\Phi-\f{1}{4}g_{\mu\nu}\nabla^\alpha\Phi\nabla_\alpha\Phi.
\ee
One can note that this corresponds to a perfect fluid with
\be\label{scalar perfect fluid}
T_{\mu\nu}=(p+\rho)U_\mu U_\nu+p\, g_{\mu\nu},
\q\q
p=\rho=-\f{1}{4}(\nabla\Phi)^2,
\q\q
U_\mu=\f{\nabla_\mu\Phi}{\sqrt{-(\nabla\Phi)^2}},
\ee
where in particular we have $w=k=1$. The homogeneous and isotropic background solution to the equations of motion \eqref{eq. EHscalar eoms} thus fits within the class of decelerating FLRW backgrounds displaying a null infinity.

The presymplectic potential, which appears as the boundary term in \eqref{eq. deltaL}, reads
\be\label{Einstein scalar potential}
\theta^\mu=\theta^\mu_\text{EH}+\theta^\mu_\text{scalar}=\f{1}{2}\sqrt{-g}\,\Big(g^{\alpha\beta}\delta\Gamma^\mu_{\alpha\beta}-g^{\mu\alpha}\delta\Gamma^\beta_{\alpha\beta}-\nabla^\mu\Phi\delta\Phi\Big).
\ee
Although this potential involves the scalar field, the Noether charge for diffeomorphisms in gravity minimally coupled to a scalar field is independent of the latter, and given simply by the gravitational Komar term
\be
K^{\mu\nu}_\xi=-\sqrt{-g}\,\nabla^{[\mu}\xi^{\nu]},
\ee
where anti-symmetrization is defined with a factor $1/2$. This can then be used to construct the Iyer--Wald density \cite{Iyer:1994ys}
\be\label{IW general Einstein scalar}
\slashed{\delta}k^{\mu\nu}_\xi
&=\delta K^{\mu\nu}_\xi-K^{\mu\nu}_{\delta\xi}+2\xi^{[\mu}\theta^{\nu]}\cr
&=\sqrt{-g}\left(\xi^{[\mu}\big(\nabla_\sigma h^{\nu]\sigma}-\nabla^{\nu]}h\big)-\xi_\sigma\nabla^{[\mu}h^{\nu]\sigma}-\f{1}{2}h\nabla^{[\mu}\xi^{\nu]}+h^{[\mu\sigma}\nabla_\sigma\xi^{\nu]}-\xi^{[\mu}\nabla^{\nu]}\Phi\delta\Phi\right),\quad
\ee
which now features the scalar field, and where the variations are
\be
h_{\mu\nu}\coloneqq\delta g_{\mu\nu},
\q\q
h^{\mu\nu}=g^{\mu\rho}g^{\nu\sigma}h_{\rho\sigma},
\q\q
h\coloneqq g^{\mu\nu}h_{\mu\nu}.
\ee
Notice that it is the $(ur)$ component of the Iyer--Wald expression~\eqref{IW general Einstein scalar}, evaluated in the limit~$r \to \infty$, that gives rise to the asymptotic surface charges at null infinity.

In order to proceed, we should now specify fall-off conditions, build a solution space, and study the asymptotic charges and their algebra. Beforehand, one can note that for a scalar field the homogeneous and isotropic exact solution to \eqref{eq. EHscalar eoms} in the Bondi coordinates \eqref{Pure FLRW bondi coord.} is given by
\be\label{FLRW-scalar backg Bondi coord.}
\beta=0,
\q\q
V=-r,
\q\q
U=0,
\q\q
\Phi=2\ln(r+u),
\ee
where we also recall that it corresponds to an equation of state with $k = 1$. Our goal is to describe perturbations on top of this background solution satisfying the boundary conditions of section~\ref{sec:asymptotically-FLRW}. We shall first consider boundary conditions leading only to supertranslations, and to trivial superrotations in the sense of spatial rotations which are Killing. Then we will extend the construction to fall-offs leading to non-trivial superrotations.

\subsection{Supertranslations}
\label{subsec: scalar_stranslations}

We begin our analysis by assuming that, on top of the background solution, the scalar field has an asymptotic expansion of the form
\be\label{fall-offs Phi no Psi0}
\Phi(u,r,\phi)=2\ln(r+u)+\sum_{n=0}^\infty\f{\Phi_n(u,\phi)}{r^n},
\q\q
\Phi_0(u,\phi)=0.
\ee
This expansion has two noteworthy features: it has no logarithmic terms at subleading orders and no term of $\O(1)$ in the radial expansion.\footnote{Since $k$ is integer, in this case it is not possible to distinguish the expansions in powers of $r$ and $a$ that we discussed in subsection~\ref{sec:Bondi hierarchy}. Still, as we shall see below, the mechanism that suggested to introduce an expansion in powers of $a$ has an effect on the identification of mass and angular momentum aspects also in this setup.} When constructing the solution space, we shall see that imposing $\Phi_0 = 0$ leads to radial fall-offs for the metric functions that are stronger than those in~\eqref{fall-offs beta V u}, with the consequence that only supertranslations survive as residual asymptotic symmetries.

\subsubsection{Solution space}

We now present the structure of the solution space obtained by solving the equations of motion~\eqref{eq. EHscalar eoms} following the steps presented in section \ref{sec:Bondi hierarchy}. It is convenient to denote the first non-vanishing term in \eqref{fall-offs Phi no Psi0} by
\be
\Phi_1(u,\phi)\eqqcolon C(u,\phi),
\ee
as this quantity will play the role of scalar ``shear'', encoding radiation. In particular, its retarded time dependence is left completely unconstrained by the field equations. On the other hand, all the subleading terms in \eqref{fall-offs Phi no Psi0} for $n\geq2$ satisfy time evolution constraints arising at each order from the radial expansion of the Klein--Gordon equation. Schematically, we have\footnote{This comes from the fact that the Klein--Gordon operator takes the form
\be
\Box\Phi=\f{1}{\sqrt{-g}}\partial_\mu\big(\sqrt{-g}\,g^{\mu\nu}\partial_\nu\Phi\big)=-\f{e^{-2\beta}}{ra^3}\big(\partial_r(ra)+2ra\partial_r\big)\dot{\Phi}+(\,\dots),
\ee
where the remaining terms in $(\,\dots)$ contain no $u$ derivative. In particular, this expression contains no time derivative of the metric functions since for the scalar field we have $\sqrt{-g}\,g^{ur}=-ar$. When expanding $\Phi$ as in \eqref{fall-offs Phi no Psi0}, the operator acting on $\dot{\Phi}$ is exactly such that the entire contribution at order $\O(r^{-5})$ becomes trivial and there is no constraint on $C$. At subleading orders the equation can then be solved as in \eqref{scalar matter EOM no Psi0}.}
\be\label{scalar matter EOM no Psi0}
\Box\Phi=\O(r^{-(n+4)})\q\Rightarrow\q\dot{\Phi}_{n\geq2}=(\,\dots),
\ee
where the dot denotes $\partial_u$. In particular, the first constraint, together with the requirement of absence of logarithmic terms\footnote{Imposing $\Box\Phi=\O(r^{-6})$ leads to $2\dot{\Phi}_2(u,\phi)=-(C+u\dot{C})$, which can be integrated up to an integration constant~$\psi_2(\phi)$. If this function is kept arbitrary, a first logarithmic term involving $\psi_2'$ is needed in $U$ at order $\O(r^{-3}\ln r)$, and logarithmic terms then propagate to the entire solution space. Here, we have chosen to set $\psi_2=0$ in order to avoid this complication, as this does not change the key features of the solution space which we want to illustrate. It is interesting to note that this structure is analogous to the four-dimensional asymptotically-flat case, where the subleading term after the shear in the expansion of the transverse metric can also contain a trace-free part sourcing logarithmic terms in the metric \cite{1985FoPh...15..605W,Chrusciel:1993hx,Geiller:2024ryw}. This trace-free part is usually set to zero in order to preserve the peeling property.\label{log footnote}} in the solution space, leads to
\be\label{Psi2}
\Phi_2(u,\phi)=-\f{u}{2} \, C(u,\phi).
\ee
With this hypothesis, the hypersurface equations \eqref{eq. Err}, \eqref{eq. Erphi} and \eqref{eq. Eur} are then solved by
\bsub\label{scalar no Psi0 solution space}
\be
\beta&=\f{1}{2r}C-\f{1}{16r^2}C(6u+C)+\O(r^{-3}),\\
U&=\f{1}{r^2}C'+\f{1}{r^3}P+\O(r^{-4}),\\
V&=-r+2M
+\O(r^{-1}),
\ee
\esub
where $P(u,\phi)$ and $M(u,\phi)$ are the two radial integration constants, which will play the respective roles of angular momentum and mass aspects. Notice that, as already announced above, the solution~\eqref{scalar no Psi0 solution space} leads to fall-off conditions that are stronger than those in~\eqref{fall-offs beta V u} as a consequence of the condition $\Phi_0=0$.

It is important to emphasize that the integration constants $P$ and $M$ do not appear at the same radial order as in the vacuum asymptotically flat case, as one might naively (and incorrectly) infer from the line element~\eqref{metric ansatz} along the lines of~\cite{Enriquez-Rojo:2020miw,Enriquez-Rojo:2022onp,Enriquez-Rojo:2022ntu}. We will shortly see that our proposal for the identification of the mass and angular momentum aspects in~\eqref{scalar no Psi0 solution space} is also consistent with the computation of the asymptotic charges, where they play the role of Coulombic data. In the same spirit, let us further stress that this change of radial order for the Bondi-like aspects by one power of $r$ is a peculiarity of the scalar-matter coupling, fixing the constant $k = 1$. For arbitrary values of~$k$, we have already observed in subsection~\ref{sec:Bondi hierarchy} that these data must arise as integration constants at orders $1/(a r^{2})$ and $r/a$ for $P$ and $M$, respectively.

We are left with the evolution equations $E_{uu}=0$ and $E_{u\phi}=0$, which, in agreement with the discussion in subsection~\ref{sec:Bondi hierarchy}, reveal at order $ (a r)^{-1}\sim r^{-2}$ the flux-balance laws for the mass and angular momentum. These equations are
\bsub\label{scalar no Psi0 flux-balance}
\be
\dot{M}&=-\f{1}{4}\big(N^2+2N+N''\big),\label{mass scalar no Psi0 flux-balance}\\
\dot{P}&=-\f{1}{12}\big(8M'-C'(1+7N)+3(u-C)N'\big),
\ee
\esub
where we have introduced the scalar news $N\coloneqq\dot{C}$. This concludes the analysis of the solution space. In summary, the hypersurface equations fix the expansions \eqref{scalar no Psi0 solution space} of the metric components in terms of the $\Phi_n$ defined in \eqref{fall-offs Phi no Psi0} as well as of $P$ and $M$, the Einstein equations $E_{uu}=0$ and~$E_{u\phi}=0$ determine the flux-balance laws \eqref{scalar no Psi0 flux-balance}, and the matter field equation fixes the $\Phi_n$ in terms of $P$, $M$, $\Phi_1=C$, as well as of an infinite tower of angle-dependent integration constants. As proven in subsection~\ref{sec:Bondi hierarchy}, once~\eqref{scalar matter EOM no Psi0} are imposed, the Einstein equations $E_{uu}=0$ and $E_{u\phi}=0$ are identically satisfied at all subleading orders. In~\eqref{Psi2} and \eqref{scalar no Psi0 solution space} we displayed explicitly only the orders that are relevant to the computation of the asymptotic charges, but following this strategy one can extend the results to the subleading orders.

Finally, we close this subsection with a discussion of the radial part of the presymplectic potential~\eqref{Einstein scalar potential}, which we find to be of the form
\be\label{scalar no Psi0 potential}
\theta^r=\f{r}{2}\delta N+\f{1}{8}\delta\big(24M+5C+(u-C)N+2C''\big)-\f{1}{2}N\delta C+\O(r^{-1}).
\ee
Disregarding total variations, which are irrelevant when computing the symplectic structure, we find that the radial presymplectic potential has a finite part given by the term $N\delta C$. One can also show that this term comes solely from the presymplectic potential $\theta^r_\text{scalar}$ of the scalar field, while the gravitational part $\theta^r_\text{EH}$ only gives the $\delta$-exact contributions. The term $N\delta C$ is the three-dimensional scalar analogue of the Ashtekar--Streubel term $N^{ab}\delta C_{ab}$ \cite{Ashtekar:1981bq}, which appears in the four-dimensional presymplectic potential for pure gravity at null infinity. Its presence indicates that the phase space is indeed radiative. One can note that the same term $N\delta C$ also appears in the study of asymptotically-flat three-dimensional Einstein--Maxwell theory \cite{Bosma:2023sxn}, which is to be expected since three-dimensional Maxwell theory is dual to a scalar field (see also \cite{Anabalon:2023efw, Cotler:2024cia} for related studies of scalars coupled to gravity in three dimensions).

\subsubsection{Residual symmetries}

We now turn to the construction of the vector fields generating the relevant residual symmetries. This was already presented in general in subsection~\ref{sec:gauge and fall-offs}, but should now be specialized to the solution space constructed above. In particular, the choice $\Phi_0=0$ imposes the additional condition
\be\label{restriction on xi from Psi0=0}
\pounds_\xi\Phi=\O(r^{-1})\q\Rightarrow\q Y'=0\q\Leftrightarrow\q Y(\phi)=Y_0,
\ee
where $Y_0$ is a constant parameter. As a result, superrotations are excluded from the residual symmetries. With this restriction, the asymptotic diffeomorphisms of the above solution space take the form
\be
\xi^u=T,\q\q
\xi^r=\f{1}{2}(T''-T)+\O(r^{-1}),\q\q
\xi^\phi=Y_0-\f{T'}{r}+\O(r^{-2}),
\ee
where $T=T(\phi)$ parametrizes the supertranslations, and where we have only written the orders which are relevant for the computation of the asymptotic charge.

We can also study the action of these diffeomorphisms on the relevant boundary data, namely the mass, the angular momentum, the scalar shear and its news. We find the transformation laws
\bsub\label{delta_scalar}
\be
\delta_\xi M&=(T\partial_u+Y_0\partial_\phi)M-\f{1}{4}\big(T+2T'N'+T''(1+N)\big),\\
\delta_\xi P&=(T\partial_u+Y_0\partial_\phi)P+\f{1}{4}\Big(3(T''C)'+uT'''+2TC'+T'\big(u-8M+C+(C-u)N\big)\Big),\\[2pt]
\delta_\xi C&=(T\partial_u+Y_0\partial_\phi)C+T+T'',\\[7pt]
\delta_\xi N&=(T\partial_u+Y_0\partial_\phi)N.
\ee
\esub
First, one should note that the scalar news $N=\dot{C}$ transforms homogeneously, so that setting $N=0$ as a non-radiative criterion is a condition which is invariant under the action of the asymptotic symmetries. We will see below that this property is due to the absence of superrotations, and that an enhanced definition of the news function is required in the presence of non-trivial superrotations. Second, one can note that the transformation laws for the mass and angular momentum do not take the same form as in the vacuum asymptotically-flat case \cite{Barnich:2006av, Oblak:2016eij}. In particular, the terms in $Y'''$ and $T'''$ appearing respectively in the transformation of the mass and of the angular momentum, and which are related to the central charge of $\mathfrak{bms}_3$, are here modified. The standard results for vacuum asymptotically-flat spacetimes are recalled in \eqref{vacuum flat transformation laws}. Here the term in $Y'''$ is absent since here there are no superrotations, and the term in $T'''$ is multiplied by a factor of $(u+3C)$. These peculiar features can directly be traced back to the fact that the mass and angular momentum do not enter the metric at the same order as in the vacuum case. It is also important to note that these transformation laws are different from what would have been obtained if we had proceeded in analogy with~\cite{Enriquez-Rojo:2022onp}, i.e. by identifying the mass and angular momentum aspects at the same orders in $U$ and $V$ as in the asymptotically-flat case. Finally, one can observe that setting $(C,P,M)=0$ in order to obtain the background solution imposes the constraint $T+T''=0$. The two solutions to this equation, together with $Y_0$, consistently reproduce the exact Killing vectors~\eqref{exact Killing FLRW Bondi} of the background.

\subsubsection{Charge algebra}
\label{Subsec. Charge algebra}

We now compute the asymptotic charges and their algebra. As is well known~\cite{Barnich:2011mi}, we shall see that the presence of radiative degrees of freedom implies the need for a modification of the standard Poisson bracket. The asymptotic surface charges are computed on a cut of $\mathcal{I}^+$ by evaluating the~$(ur)$ component of the Iyer--Wald density~\eqref{IW general Einstein scalar} in the limit $r \to \infty$, and integrating the result over the celestial circle~$S^1$. We find that the resulting charges are finite and given by
\be\label{full charge scalar no Psi0}
\slashed{\delta}Q_\xi^\text{finite}=\delta Q_\xi+\Xi_\xi[\delta],
\q
Q_\xi=\f{1}{2}\oint \Big(T(4M+C)-3Y_0P \Big),
\q
\Xi_\xi[\delta]=\f{1}{2}\oint TN\delta C.
\ee
Here we have written the result as the sum of an integrable part and a flux term. While this split is notoriously ambiguous, it can be justified by the following observations. First, the integrable part is conserved when $N=0$, i.e.\ in the radiative vacuum where the scalar news vanishes. When this vacuum condition is enforced, one can use the flux-balance laws \eqref{scalar no Psi0 flux-balance} to obtain
\be
\dot{Q}_\xi\big|_{N=0}=0.
\ee
Second, one can check that the integrable part of the charge arises from the flux of a Wald--Zoupas symplectic potential identified in \eqref{scalar no Psi0 potential} as $\theta_\text{WZ}=-N\delta C \,\de u\wedge \de \phi$. This potential satisfies the requirements of vanishing in the absence of news, and of being invariant under the action of the anomaly operator $\Delta_\xi\coloneqq\delta_\xi-L_\xi$ introduced in \cite{Freidel:2021cjp}, where $\delta_\xi$ denotes the variation induced by the transformations \eqref{delta_scalar}, while $L_\xi$ is the Lie derivative defined with respect to the boundary vector field \eqref{residual diffeo on scri}. A non-vanishing action of this operator would signal a lack of covariance of the construction and in our case we get
\be\label{covariant WZ}
\theta_\text{WZ}\big|_{N=0}=0,
\q\q
\Delta_\xi\big(\theta_\text{WZ}\big)\big|_{\mathcal{I}^+} = 0,
\q\q
\oint I_\xi\theta_\text{WZ}=-\f{1}{2}\oint N\delta_\xi C=\dot{Q}_\xi,
\ee
where we evaluated $\Delta_\xi$ on the pullback at $\mathcal{I}^+$ of the symplectic potential and the last equality requires to use \eqref{scalar no Psi0 flux-balance}.

Since the full charge \eqref{full charge scalar no Psi0} is non-integrable, the computation of the algebra of the integrable charges can be done using the Barnich--Troessaert bracket \cite{Barnich:2011mi}. This bracket is defined as
\be
\big\{Q_{\xi_1},Q_{\xi_2}\big\}_\text{BT}\coloneqq\delta_{\xi_2}Q_{\xi_1}+\Xi_{\xi_2}[\delta_{\xi_1}],
\ee
and has the property of reproducing the algebra of residual symmetries up to a possible 2-cocycle~$\K$. The Barnich--Troessaert bracket generically gives an expression of the form
\be\label{BT bracket}
\big\{Q_{\xi_1},Q_{\xi_2}\big\}_\text{BT}=Q_{[\xi_1,\xi_2]_*}+\K_{\xi_1,\xi_2},
\ee
where the modified Lie bracket taking into account the field-dependence of the vector fields is
\be
\big[\xi_1,\xi_2\big]_*\coloneqq\big[\xi_1,\xi_2\big]-\delta_{\xi_1}\xi_2+\delta_{\xi_2}\xi_1.
\ee
Using the split \eqref{full charge scalar no Psi0}, one can indeed show that the bracket of the integrable charges gives
\be
\big\{Q_{\xi_1},Q_{\xi_2}\big\}_\text{BT}=Q_{\xi_{12}},
\q\q
\xi_{12}=\big(Y_{0,1}T_2'-Y_{0,2}T_1'\big)\partial_u+0\times\partial_\phi,
\ee
which shows in particular that the 2-cocycle is vanishing, i.e. $\K=0$.

\subsection{Superrotations}
\label{sec:scalar_superrotations}

On top of the background solution, we now assume that the scalar field admits the asymptotic expansion
\be\label{fall-offs Phi with Psi0}
\Phi(u,r,\phi)=2\ln(r+u)+\sum_{n=0}^\infty\f{\Phi_n(u,\phi)}{r^n},
\q\q
\Phi_0(u,\phi)\neq0.
\ee
This should be contrasted with the choice~\eqref{fall-offs Phi no Psi0} made above, where $\Phi_0$ was set to zero. The non-vanishing function $\Phi_0 \neq 0$ precisely leads to the fall-off conditions~\eqref{fall-offs beta V u} and, as a consequence, allows for non-trivial superrotations.

\subsubsection{Solution space}
\label{subsec. superrot-solution space}

We present the structure of the solution space for $\Phi_0\neq0$, once again following the algorithm outlined in subsection~\ref{sec:Bondi hierarchy}. In addition to~\eqref{scalar matter EOM no Psi0}, the Klein--Gordon equation imposes that at leading order the scalar field only depends on the angular coordinate,
\be
\Phi_0(u,\phi)=\psi(\phi),
\ee
and this new function entering the solution space is responsible for the appearance of non-trivial superrotations. The subleading term $\Phi_1$ is still left unconstrained and is identified with the scalar shear, $\Phi_1(u,\phi)\eqqcolon C(u,\phi)$. The Klein--Gordon equation, together with the requirement that no logarithmic terms appear in the solution,\footnote{The discussion closely parallels that of footnote~\ref{log footnote}, with the difference that the first logarithmic contribution now arises at order $\mathcal{O}(r^{-3}\ln r)$ and is sourced by $\psi_2' + \psi_2 \psi'$. These logarithmic terms can be removed by choosing $\psi_2 = A e^{-\psi}$. For simplicity, we set $A = 0$, and hence $\psi_2 = 0$, since this choice does not affect the subsequent results for the charges and their algebra.} once again leads to the expression for $\Phi_2$ given in~\eqref{Psi2}. The hypersurface equations are then solved by the following radial expansions:
\bsub\label{scalar with Psi0 solution space}
\be
\beta&=\f{1}{2r}C-\f{1}{16r^2}C(6u+C)+\O(r^{-3}),\\
U&=-\f{1}{r}\psi'+\f{1}{r^2}\big(C'+u\psi'\big)+\f{1}{r^3}P+\O(r^{-4}),\\
V&=r\left(-1+\f{1}{2}(\psi')^2+\f{3}{2}\psi''\right)+2M
+\O(r^{-1}).
\ee
\esub
Due to the presence of $\psi$, the resulting fall-off conditions now coincide with those anticipated in~\eqref{fall-offs beta V u}, while we find the same radial integration constants as in the previous subsection.

The evolution equations encoded at order $r^{-2}$ in $E_{uu} = 0$ and $E_{u\phi} = 0$ are now more involved than in~\eqref{scalar no Psi0 flux-balance}:
\bsub\label{pre-flux-balance}
\be
\begin{split}
\dot{M}
&=-\f{1}{4}\big(N^2+2N+N''\big)\cr
&\pe-\f{1}{8}\Big(3(\psi'')^2+3\psi'\psi'''+3\psi''''-9(\psi')^2\psi''+4(1+N)\big(\psi''-(\psi')^2\big)\Big),
\end{split}\\
\begin{split}
\dot{P}
&=-\f{1}{12}\big(8M'-C'(1+7N)+3(u-C)N'\big)+\f{1}{12}\psi'\big(16M-13u+4C''\big)\cr
&\pe-\f{1}{12}(u+C)\Big(2(\psi')^3+6\psi'''-2\psi''\psi'-5\psi'(1+N)\Big)+\f{1}{3}C'\big(\psi''-(\psi')^2\big).\q
\end{split}
\ee
\esub
This completes the analysis of the enlarged solution space, to a radial order sufficient for our purposes. Solving the field equations at subleading order determines the remaining subleading terms in~\eqref{scalar with Psi0 solution space}, as well as the evolution equations for the higher coefficients~$\Phi_n$. It is important to stress that evolution equations \eqref{pre-flux-balance} cannot immediately be interpreted as ``physical'' flux-balance laws since, as we shall discuss in subsection~\ref{sec:transformations with psi}, when the field $\psi$ is added to the solution space the condition $N = 0$ is not invariant anymore under the action of asymptotic symmetries and, as such, cannot characterize the absence of scalar radiation.

Finally, the radial component of the presymplectic potential is also more involved than in~\eqref{scalar no Psi0 potential}, and reads
\be\label{eq. theta_scalar_srotations}
\theta^r
&=\f{r}{2}\Big(\delta\big(N+3(\psi')^2+5\psi''\big)-3\partial_\phi\big(\delta\psi\psi'\big)-N\delta\psi\Big)+\f{1}{4}\partial_\phi\big(C(\delta\psi\psi'+\delta\psi')\big)\\
&\pe+\f{1}{8}\delta\Big(24M+5C+(u-C)N+2C''+(u-2C)\big(3(\psi')^2+4\psi''\big)+5C(\psi')^2-2C'\psi'\Big)\cr
&\pe-\f{1}{2}\left(N+\f{5}{4}(\psi')^2+\f{9}{2}\psi''\right)\delta C+\f{1}{4}\delta\psi\Big((u+C)(\psi')^2+C(2\psi''-1)-uN-8M\Big)+\O(r^{-1}).\quad\nn
\ee
Several features are worth emphasizing. First, this radial symplectic potential contains an $\O(r)$ divergent contribution that is neither a total variation nor a total angular derivative. This divergence will therefore manifest itself at the level of the asymptotic charges through the appearance of a divergent term. However, since $\psi$ does not depend on $u$, the divergent contribution to the radial potential is a corner contribution that can be written as a total $u$ derivative. Following \cite{Compere:2008us,Freidel:2019ohg,McNees:2023tus,Campoleoni:2023eqp,Riello:2024uvs,Campoleoni:2025bhn,Ammon:2025avo}, one can then exploit the corner ambiguity in the definition of the symplectic structure in order to renormalize the charges and thereby discard the divergent contribution.

Second, at order $\O(1)$, after discarding irrelevant total variations and total angular derivatives, new contributions appear in addition to the familiar Ashtekar--Streubel term $N\delta C$. These additional terms arise due to the presence of $\psi \neq 0$ and correspond to the third line in the expression~\eqref{eq. theta_scalar_srotations}. At this stage, these new terms cannot be directly interpreted as originating from scalar radiation, since it is not clear which physical notion of radiation should replace the news $N = \dot{C}$. This situation is closely analogous to what happens in four-dimensional asymptotically flat spacetimes when relaxing the boundary conditions so as to extend the historical BMS group to the generalized BMS group, with an arbitrary sphere metric and arbitrary $\mathrm{Diff}(S^2)$ symmetries. In that case as well, the presymplectic potential acquires additional contributions beyond the standard $N^{ab} \delta C_{ab}$ term~\cite{Compere:2018ylh}.

\subsubsection{Residual symmetries}
\label{sec:transformations with psi}

Since the solution space now satisfies the fall-off conditions~\eqref{fall-offs beta V u}, the construction of the residual symmetries presented in subsection~\ref{sec:gauge and fall-offs} applies straightforwardly, and we recover supertranslations together with non-trivial superrotations. Using the expansions~\eqref{fall-offs Phi with Psi0} and~\eqref{scalar with Psi0 solution space}, we find that the asymptotic Killing vectors take the form
\be\label{bmsk=1 diffeos}
\xi^u=f,\quad\;\;\;
\xi^r=-\f{r}{2}Y'+\f{1}{4}\left(f''-f-f'\psi'-\f{u}{2}Y'\right)+\O(r^{-1}),\quad\;\;\;
\xi^\phi=Y-\f{f'}{r}+\O(r^{-2}),
\ee
with
\be
f=T+\f{3}{2}uY'.
\ee
The symmetries are therefore parametrized by supertranslations $T(\phi)$ and superrotations $Y(\phi)$.

We can study the action of these diffeomorphisms on the relevant boundary data, namely the radial integration constants $M$ and $P$, the scalar shear, its time derivative, and the field $\psi$. We find the following transformation laws:
\bsub\label{scalar transformations with Y}
\be
\label{deltaM}
\delta_\xi M
&=\left(f\partial_u+Y\partial_\phi+\f{5}{2}Y'\right)M-\f{1}{4}\big(f+2f'N'+f''(1+N)\big)\\
&\pe-\f{1}{4}f'\big((\psi')^3+7\psi'\psi''+3\psi'''\big)-\f{1}{2}f''(\psi')^2+\f{1}{2}f'''\psi'-\f{5}{8}uY'-\f{1}{2}Y''C'-\f{3}{8}(u+C)Y''',\nn\\
\label{deltaP}
\delta_\xi P &=\big(f\partial_u+Y\partial_\phi+2Y'\big)P+\f{1}{4}\Big(3(f''C)'+uf'''+2fC'+f'\big(u-8M+C+(C-u)N\big)\Big)\nn\\
&\pe+\f{3}{4}fu\psi'-\f{1}{2}f'\big((u+C)(\psi')^2+(u+3C)\psi''+C'\psi'\big)+\f{1}{4}f''(u-3C)\psi'\nn\\
&\pe+\f{1}{8}u(3u\psi'+2C')Y'+\f{1}{8}u(u-2C)Y'',\\
\delta_\xi C&=\left(f\partial_u+Y\partial_\phi+\f{1}{2}Y'\right)C+f+f''-2f'\psi'+\f{1}{2}uY',\label{transformation C with psi}\\
\delta_\xi N&=\big(f\partial_u+Y\partial_\phi+2Y'\big)N+2Y'-3Y''\psi'+\f{3}{2}Y''',\\[5pt]
\delta_\xi\psi&=Y\psi'-Y'.\label{transformation psi}
\ee
\esub
One immediately sees that the presence of $\psi \neq 0$ generates many additional, non-homogeneous terms in the transformation laws. The field $\psi$ transforms according to~\eqref{transformation psi}, which clearly shows that it is responsible for the appearance of non-trivial superrotations. The non-homogeneous transformation of the news $N$ leads to an ambiguity in the definition of a non-radiative vacuum, as anticipated, since the condition $N=0$ is no longer invariant under residual diffeomorphisms.

\subsubsection{Vacuum structure and covariant news}
\label{sec:vacuum}

In order to formulate a notion of radiative vacuum we now need to characterize the orbit of the previous residual symmetries acting on the FLRW background. This is similar to what has been done in \cite{Compere:2016jwb,Compere:2018ylh} for four-dimensional asymptotically-flat spacetimes. To do so, we start from a parametrization of the background involving a different set of coordinates, namely the conformal ones of equation~\eqref{conformal coords}, where the metric and the scalar field are given by\footnote{Note that we have relabeled the conformal coordinates of \eqref{conformal coords} in order to keep $x^\mu=(u,r,\phi)$ for the Bondi coordinates reached after the diffeomorphism.}
\bsub\label{starting_g}
\be
\de s^2&=\tilde{g}_{\mu\nu}\de\tilde{x}^\mu\de\tilde{x}^\nu=a(\tau)^2\big(\!-\!\de\tau^2+\de\rho^2+\rho^2\de\theta^2\big),\qquad a(\tau)=\tau,\\
\Phi(\tau)&=2\ln(\tau).
\ee
\esub
We then act on this background solution with an arbitrary finite diffeomorphism generating the change of coordinates
\be
\tilde{x}^\mu=(\tau,\rho,\theta)\mapsto x^\mu=(u,r,\phi),
\ee
and we impose that the result satisfies the gauge-fixing conditions \eqref{metric ansatz} as well as the boundary conditions \eqref{fall-offs beta V u}. As detailed in appendix~\ref{Appendix vacuum}, the resulting \textit{finite} diffeomorphism is parametrized by two angular functions $\T=\T(\phi)$ and $\W=\W(\phi)$. Adopting the same terminology as in \cite{Compere:2018ylh}, these can respectively be identified as the supertranslation and the superboost fields. They transform accordingly under infinitesimal $\mathfrak{bms}_3^{k=1}$ residual symmetries as
\bsub\label{Transfo W and T}
\be
\delta_\xi\W&=Y\W'+Y',\label{Transfo W}\\
\delta_\xi\T&=\left(Y\partial_\phi-\f{3}{2}Y'\right)\T+T,
\ee
\esub
where the conformal weight $-3/2$ in the transformation law of $\T$ is characteristic of the case $k=1$.

Acting on the background solution \eqref{starting_g} in conformal coordinates with the diffeomorphism generated by $\T$ and $\W$ gives a solution in the form
\bsub\label{vacuum-valued bondi param}
\be
\beta&=\f{1}{2r}C_\text{vac}+\O(r^{-2}),\\
U&=\f{1}{r}\W'+\O(r^{-2}),\\
V&=r\left( -1+\f{1}{2}\big(\W'\big)^2-\f{3}{2}\W''\right)+\O(1),\\
\Phi&=2\ln(r+u)-\W+\O(r^{-1}),\label{vacuum Phi}
\ee
\esub
where $C_\text{vac}$ is defined in \eqref{Cvac}. Through this procedure, we have obtained a two-parameter family of metrics satisfying our definition of asymptotically-FLRW spacetimes and that must be considered as vacua. Comparing this expression with the solution space discussed in subsection~\ref{subsec. superrot-solution space} allows one to extract the vacuum values of the various boundary data appearing in the expansion. In particular, for the scalar field one finds
\bsub\label{Cvac}
\be
\psi&\stackrel{\text{vac}}{=}\psi_\text{vac}\coloneqq-\W,\\
C&\stackrel{\text{vac}}{=}C_\text{vac}\coloneqq2\T'\W'+\T''+\big(\T+u\big)\left(e^{2\W}+\f{3}{4}\left[\big(\W'\big)^2+2\W''\right]\right)-u,
\ee
\esub
while for the gravitational sector one has
\bsub
\be
M\stackrel{\text{vac}}{=}\bar{M}\big[\W,C_\text{vac}\big]&\coloneqq-\f{1}{2}\W'C_\text{vac}'+\f{1}{16}\left(5\big(\W'\big)^2-6\W''-4e^{2 \W}\right)\big(u+C_\text{vac}\big),\\[5pt]
P\stackrel{\text{vac}}{=}\,\bar{P}\big[\W,C_\text{vac}\big]&\coloneqq\f{1}{8}\Big(2\left(u+3C_\text{vac}\right)C_\text{vac}'-3C_\text{vac}\W'\big(2 u+C_\text{vac}\big)\Big),
\ee
\esub
where the symbol $\stackrel{\text{vac}}{=}$ means that the corresponding expressions are evaluated on the diffeomorphism orbit of the background. As expected from their transformation laws \eqref{transformation psi} and \eqref{Transfo W}, we observe that $\psi$ and $\W$ can be identified (up to a minus sign). We also note that $C_\text{vac}$ contains both a $u$-independent part and a term linear in $u$. This structure naturally motivates the introduction of a covariant news, which will be defined in the next paragraph.

Even in the presence of $\psi(\phi)$, the only quantity in the solution space with an unspecified dependence on $u$ remains $C=C(u,\phi)$. The radiative degrees of freedom of the scalar theory can therefore only be encoded in a combination of boundary data involving the scalar shear but vanishing on the vacua defined above. To this end, one can introduce the covariant news
\be\label{News}
\hat{N}
&\coloneqq\dot{C}-\dot{C}_\text{vac}\big|_{\W\,=\,-\psi}\cr
&\,=\dot{C}+\f{3}{2}\psi''-\f{3}{4}(\psi')^2+1-e^{-2\psi}.
\ee
Importantly, this quantity transforms covariantly as a conformal weight-$2$ tensor under a $\mathfrak{bms}_3^{k=1}$ transformation. Indeed, one can check using \eqref{transformation C with psi} and \eqref{transformation psi} that we have
\be\label{transfo News}
\delta_\xi\hat{N}=\big(f\partial_u+Y\partial_\phi+2Y'\big)\hat{N}.
\ee
Moreover, this covariant news $\hat{N}$ reduces to $N$ when $\psi$ is set to zero, as the exponential term cancels the $+1$ contribution. These properties motivate us to identify non-radiative solutions by the condition $\hat{N}=0$. Consistently, we have by construction that the value of this covariant news on the vacuum configurations is vanishing, i.e. $\hat{N}\stackrel{\text{vac}}{=}0$.

This construction is reminiscent of the introduction of the Geroch tensor in four-dimensional asymptotically flat spacetimes \cite{Geroch1977,Dray:1984rfa}. In particular, in the context of the so-called extended BMS and generalized BMS extensions including superrotations, the more recent references \cite{Compere:2018ylh,Campiglia:2020qvc,Nguyen:2022zgs,Rignon-Bret:2024wlu,Rignon-Bret:2024gcx} explain how a covariant news can be built using a Geroch tensor related to a CFT stress-energy tensor. Here we have resorted as well to a purely angular-dependent quantity in order to construct a covariant news tensor, although our three-dimensional analogue of the Geroch tensor was built in terms of the object $\psi$ already present in the solution space.

Interestingly, the fact that the covariant news \eqref{News} represents the radiative flux reaching null infinity can also be interpreted thermodynamically in terms of the local energy density produced by perturbations over the background. This can be seen by computing the energy fluctuation
\be
\Delta\text{E}(u)\coloneqq\text{E}\big|_{S_{\infty}^1}-\text{E}_\text{vac}\big|_{S_{\infty}^1}\coloneqq\lim_{r\to\infty}\left(\oint_{S^1}\de\phi\,\sqrt{-g}\,\rho-\oint_{S^1}\de\phi\,\sqrt{-g}\,\rho_\text{vac}\right)=\oint_{S_{\infty}^1}\de\phi\,\hat{N}(u,\phi),
\ee
which reproduces the covariant news $\hat{N}$. For this computation we have used $\rho$ given in \eqref{scalar perfect fluid}, and~$\rho_\text{vac}$ is the same expression evaluated on the background \eqref{vacuum-valued bondi param}.

\subsubsection{Covariant aspects and Cotton scalars}
\label{sec:Cotton}

With the help of the vacuum structure defined above, we can now define an improved notion of mass and angular momentum aspects that vanish on the two-parameter family of vacua and that transform among each other in the absence of radiation, i.e., when $\hat{N}=0$. The definition of these improved mass and angular momentum aspects is
\bsub\label{covariant M and P}
\be
\M
&\coloneqq M-\bar{M}\big[\!-\!\psi,C\big]+\f{1}{4}(C+u)\hat{N}\cr
&\,=M-\f{1}{4}\big(u+2\psi'C'\big)+\f{1}{16}(C+u)\left(4e^{-2\psi}-5(\psi')^2-6\psi''\right)+\f{1}{4}(C+u)\hat{N}\cr
&\stackrel{\text{vac}}{=}0,\\
\P
&\coloneqq P-\bar{P}\big[\!-\!\psi,C\big]\cr
&\,=P-\f{3}{8}C(C+2u)\psi'-\f{1}{4}(3C+u)C'\cr
&\stackrel{\text{vac}}{=}0,
\ee
\esub
where we have made explicit the fact that, by construction, these improved aspects vanish on vacuum configurations.

Since $\hat{N}\stackrel{\text{vac}}{=}0$, one can add arbitrary contributions involving $\hat{N}$ to the previous expressions for the improved mass and angular momentum aspects. Here this ambiguity has been fixed by adding in $\M$ a term such that the resulting transformation laws take the form
\bsub\label{covariant transformation laws}
\be
\delta_\xi\N&=\left(f\partial_u+Y\partial_\phi+\f{7}{2}Y'\right)\N,\\
\delta_\xi\J&=\left(f\partial_u+Y\partial_\phi+\f{6}{2}Y'\right)\J+f'\N,\\
\delta_\xi\M&=\left(f\partial_u+Y\partial_\phi+\f{5}{2}Y'\right)\M-f'\J,\\
\delta_\xi\P&=\left(f\partial_u+Y\partial_\phi+\f{4}{2}Y'\right)\P-2f'\M,
\ee
\esub
where we have also introduced and given the transformation laws for
\be\label{definitions curly N and J}
\N\coloneqq\f{1}{2}\partial_u\hat{N}=\f{1}{2}\ddot{C},
\q\q
\J\coloneqq\f{1}{2}\big(\hat{N}'+2\hat{N}\psi'\big).
\ee
Remarkably, these transformation laws are analogous (up to a change in the conformal weights) to those obtained respectively for the Weyl scalars $\Psi_4^0$, $\Psi_3^0$, $\Psi_2^0$, and $\Psi_1^0$ in four-dimensional asymptotically-flat spacetimes \cite{Barnich:2019vzx,Freidel:2021qpz,Geiller:2024bgf}. This begs the question of whether the covariant functionals which we have built by studying the vacuum structure can be recovered as Newman--Penrose scalars. We now show that this is indeed the case.

In three-dimensional spacetimes the Weyl tensor is identically vanishing, and the obstruction to conformal flatness is instead captured by the Cotton tensor. The latter, being symmetric and trace-free, has 5 independent components, which can conveniently be encoded into real Cotton scalars obtained by projections on a triad. For this, let us consider the vectors
\be
\ell=\f{1}{a}\partial_r,
\q\q
n=\f{1}{a}e^{-2\beta}\left(\partial_u+\f{V}{2r}\partial_r+U\partial_\phi\right),
\q\q
m=\f{1}{ar}\partial_\phi.
\ee
These are such that
\be
g^{\mu\nu}=-2\ell^{(\mu}n^{\nu)}+m^\mu m^\nu,
\q\q
\ell_\mu n^\mu=-1=-m_\mu m^\mu,
\ee
with all other contractions vanishing. Starting from the Cotton tensor
\be
C_{\mu\nu}\coloneqq\sqrt{-g}\,{\epsilon_\mu}^{\rho\sigma}\nabla_\rho\left(R_{\sigma\nu}-\f{1}{4}g_{\sigma\nu}R\right),
\ee
one can build Cotton scalars\footnote{The projection on $\ell$ and $n$ is not independent since $C_{mm}=2C_{\ell n}$.}
\be
\Psi_0\coloneqq C_{\ell\ell},
\q\quad
\Psi_1\coloneqq C_{\ell m},
\q\quad
\Psi_2\coloneqq C_{mm},
\q\quad
\Psi_3\coloneqq C_{nm},
\q\quad
\Psi_4\coloneqq C_{nn}.
\ee
We then find that their expansion is
\bsub\label{3d Cotton peeling}
\be
\Psi_0&=-\f{3\P}{r^8}+\O(r^{-9}),\label{Psi0}\\[4pt]
\Psi_1&=\f{3\M}{r^7}+\O(r^{-8}),\\[4pt]
\Psi_2&=-\f{2\J}{r^6}+\O(r^{-7}),\\[4pt]
\Psi_3&=-\f{\N}{r^5}+\O(r^{-6}),\\[4pt]
\Psi_4&=-\f{\psi'\N}{r^5}+\O(r^{-6}),
\ee
\esub
which, as anticipated, reproduces the covariant functionals defined above. For comparison between this three-dimensional (3d) construction and the properties of the four-dimensional (4d) Weyl scalars in asymptotically-flat spacetimes without matter, one can note that at leading order the analogy is
\be
\Psi_0^\text{3d}\leftrightarrow\Psi_1^\text{4d},
\q\q
\Psi_1^\text{3d}\leftrightarrow\Psi_2^\text{4d},
\q\q
\Psi_2^\text{3d}\leftrightarrow\Psi_3^\text{4d},
\q\q
\Psi_3^\text{3d}\leftrightarrow\Psi_4^\text{4d},
\ee
apart from the fall-offs which are different.\footnote{Just like in the four-dimensional case, introducing logarithmic terms in the solution space (which is the case if we keep $\psi_2(\phi)\neq0$ as discussed in footnote \ref{log footnote}), leads to modifications of the fall-off conditions \eqref{3d Cotton peeling} which can be interpreted as violations of the (here three-dimensional) peeling. In this case, we find instead of \eqref{Psi0} that $\Psi_0$ behaves as $\Psi_0=\O\big(\!\log(r)r^{-8}\big)$, while the fall-offs of the other Cotton scalars are unaffected.} In the 4d case, $\Psi_0^\text{4d}$ captures the so-called incoming radiation, namely the degrees of freedom encoded in the angular metric $g_{ab}$ at order $\O(r^{-1})$ and beyond. In the present case, this role is instead played by the subleading terms~$\Phi_{n\geq 3}$ in the expansion~\eqref{fall-offs Phi with Psi0} of the scalar field. However, these contributions do not appear to be encoded in~$\Psi_4^\text{3d}$, so that no direct analogue of the Weyl scalar~$\Psi_0^\text{4d}$ arises from this Cotton scalar. It is also worth noting that, although the Cotton scalars capture the covariant aspects identified through the analysis of the vacuum structure, they do not coincide with the variables entering the three-dimensional Newman--Penrose formalism~\cite{Hall:1987vw,Aliev:1995cf,Milson:2012ry,Barnich:2016lyg}. The latter is instead formulated entirely in terms of spin coefficients and projections of the Ricci tensor. In subsection~\ref{sec:subleading}, we further develop the analogy with the four-dimensional case by rewriting the evolution equations in terms of the covariant aspects that appear at leading order in the Cotton scalars.

\subsubsection{Charge algebra}
\label{sec:charge algebra with Y}

We can finally turn to the study of the asymptotic charges obtained from the $(ur)$ component of~\eqref{IW general Einstein scalar}. Due to the non-trivial superrotations, these charges now contain divergent contributions. The first one is of order $\mathcal{O}(r^2)$, but it is a total angular derivative and therefore vanishes upon integration over the celestial circle. The second one appears at order $\mathcal{O}(r)$. The final result is given in equation \eqref{Iyer-Wald sc} of appendix \ref{app:bare charges}, where we have made an arbitrary split of the finite $\mathcal{O}(1)$ contribution into an integrable part and a flux term. With the split given in \eqref{bare split IW scalar Y}, the resulting cocycle is field-dependent and given by \eqref{cocycle IW scalar Y}. Since this split is arbitrary, one can of course rewrite
\be\label{split ambiguity}
\delta Q_\xi+\Xi_\xi[\delta]=\delta\big(Q_\xi+S_\xi\big)+\Xi_\xi[\delta]-\delta S_\xi\eqqcolon\delta\tilde{Q}_\xi+\tilde{\Xi}_\xi[\delta],
\ee
and thereby redefine at will the integrable and the flux parts with a shift $S_\xi$. Under such a shift, the Barnich--Troessaert bracket~\eqref{BT bracket} is not invariant and becomes
\be\label{BT bracket new split}
\big\{\tilde{Q}_{\xi_1},\tilde{Q}_{\xi_2}\big\}_\text{BT}=\tilde{Q}_{[\xi_1,\xi_2]_*}+\tilde{\K}_{\xi_1,\xi_2}
\q\q
\tilde{\K}_{\xi_1,\xi_2}\coloneqq \K_{\xi_1,\xi_2}-\big(\delta_{\xi_1}S_{\xi_2}-\delta_{\xi_2}S_{\xi_1}+S_{[\xi_1,\xi_2]_*}\big),
\ee
where the last term gives the cocycle in the new split in terms of the cocycle in the initial split and the contribution of the shift.

The split ambiguity manifests itself whenever the charge features a non-integrable contribution. In the absence of superrotations and for $\psi = 0$, the decomposition adopted in~\eqref{full charge scalar no Psi0} is preferred, as the flux vanishes in the absence of radiation and the integrable part of the charge is therefore conserved. In this situation, the cocycle was also found to vanish. A natural question is whether similar results can be obtained for the charges \eqref{bare split IW scalar Y} in the presence of superrotations and $\psi\neq0$. This is far from being obvious since now the charge contains many non-integrable contributions which cannot easily be written in terms of the covariant news $\hat{N}$. Similar complications appear in four-dimensional asymptotically-flat spacetimes when considering a generic sphere as the celestial metric, which corresponds to the so-called generalized BMS boundary conditions.

Inspired by the Wald--Zoupas construction \cite{Wald:1999wa} presented in \cite{Compere:2018ylh} and \cite{Fiorucci:2021pha} (see section~5.3.2), one can choose a split of the charges such that the integrable part $Q_\xi$ satisfies the following two conditions:
\be\nn
&\text{1) the integrable part vanishes on vacuum configurations, i.e. }Q_\xi\stackrel{\text{vac}}{=}0,\cr
&\text{2) the integrable part is conserved in the absence of radiation, i.e. }\dot{Q}_\xi\big|_{\hat{N}=0}=0.
\ee
We note that these conditions still leave an ambiguity in \eqref{split ambiguity} of the form
\be
\delta Q_\xi+\Xi_\xi[\delta]=\delta\big(Q_\xi+F[\hat{N}]\big)+\Xi_\xi[\delta]-\delta F[\hat{N}],
\ee
which amounts to redefining the integrable part up to an arbitrary functional of the covariant news~$\hat{N}$. This ambiguity could potentially be fixed by requiring, as in \cite{Compere:2018ylh,Donnay:2021wrk,Donnay:2022hkf}, that the soft and hard parts of the flux separately represent the $\mathfrak{bms}_3^k$ algebra, and/or by studying the covariance properties of the Wald--Zoupas potential as in \cite{Grant:2021sxk,Odak:2022ndm,Rignon-Bret:2024wlu,Rignon-Bret:2024gcx}. We leave these more detailed investigation for future work, and simply write here a proposal satisfying the above criteria 1) and~2). For the integrable part this is
\be\label{scalar WZ charge with Y}
Q_\xi=\f{1}{2}\oint\Big(4T\M-3Y\P+2uY\big(2\M\psi'-\M'\big)\Big),
\ee
and the non-integrable part can be reconstructed by comparison with \eqref{bare split IW scalar Y}. Since the integrable part is now written in terms of the covariant functionals introduced in \eqref{covariant M and P}, it is clear that we have~$Q_\xi\stackrel{\text{vac}}{=}0$. In order to check the second criterion, we can use the rewriting of the evolution equations \eqref{pre-flux-balance} given below in \eqref{higher spin EOM}. This leads to
\be
\dot{Q}_\xi=\f{1}{2}\oint\Big(\big(2T-uY\partial_\phi+2uY\psi'\big)\big((C+u)\N-\J'\big)+2Y(C+u)\J\Big).
\ee
Since this flux depends only on $\J$ and $\N$ defined in \eqref{definitions curly N and J} through $\hat{N}$, we immediately get that
\be
\dot{Q}_\xi\big|_{\hat{N}=0}=0,
\ee
so that the integrable part of the split is conserved in the absence of radiation.

Starting from the integrable charge \eqref{scalar WZ charge with Y}, one can compute the shift with respect to the split given in \eqref{bare split IW scalar Y}, and then use \eqref{BT bracket new split} to find the new cocycle produced by the Barnich--Troessaert bracket of the charges \eqref{scalar WZ charge with Y}. This is a lengthy computation which we do not reproduce here, but unfortunately the resulting 2-cocycle is once again field-dependent. This suggests that extra criteria should be leveraged in order to further improve the choice of integrable part \eqref{scalar WZ charge with Y}, such as the relationship between the charges and a covariant choice of Wald--Zoupas symplectic potential. In the four-dimensional asymptotically-flat case this was done for the extended BMS group in \cite{Odak:2022ndm,Rignon-Bret:2024wlu,Rignon-Bret:2024gcx}. We leave the question of adapting this construction to the present context for future work.

\subsection{Subleading structure}
\label{sec:subleading}

We conclude this section on the coupling to scalar matter by discussing the evolution equations rewritten in terms of the covariant aspects introduced above, which appear at leading order in the Cotton scalars. We have indeed seen that these quantities provide the three-dimensional analogue of the leading Weyl scalars in four-dimensional asymptotically flat spacetimes, and they play a central role in the definition of the Wald--Zoupas charges. An interesting structure also appears when they are used to rewrite the evolution equations, and this reveals a striking resemblance with the so-called higher spin charges which have been introduced in four-dimensional asymptotically-flat spacetimes \cite{Freidel:2021dfs,Freidel:2021ytz,Compere:2022zdz,Geiller:2024bgf,Cresto:2024fhd,Cresto:2024mne}. After establishing this analogy, we explain how the spin-3 charge is related, as in the four-dimensional case, to the Newman--Penrose charges~\cite{Newman:1965ik,Newman:1968uj}.

\subsubsection{Covariant evolution equations}

For $s\in\{-2,-1,0,1,2\}$, let us consider the charge aspects $Q_s$ defined in terms of the leading Cotton scalars and in terms of $\Phi_3$ as
\bsub\label{spin s charges}
\be
Q_{-2}&\coloneqq\N,\\
Q_{-1}&\coloneqq\J,\\
Q_0&\coloneqq-2\M,\\
Q_1&\coloneqq3\P,\\
Q_2&\coloneqq12\Phi_3+\f{1}{2}C\big(C^2+3uC-6u^2\big).\label{3d spin-2 charge}
\ee
\esub
These charges are all such that $Q_s\stackrel{\text{vac}}{=}0$. This is manifest for $-2\leq s\leq1$ as a consequence of the properties of the vacuum structure presented above. For $s=2$, the result follows from analyzing the subleading terms in \eqref{vacuum Phi}. Let us now define the derivative operator
\be\label{covariant D phi}
\D_\phi\coloneqq\partial_\phi-2(s+1)\psi',
\ee
which is understood as acting on the charges $Q_s$. This defines a covariant derivative with respect to the superrotations. Indeed, given a functional $G$ which transforms under superrotations as $\delta_YG=\big(Y\partial_\phi-2(s+1)Y'\big)G$, then we have that $\delta_Y(\D_\phi G)=\big(Y\partial_\phi-2(s+1)Y'\big)(\D_\phi G)$ by virtue of the transformation law \eqref{transformation psi}.

Using these quantities, a remarkable rewriting of the evolution equations and of the transformation laws occurs, and we arrive at the compact and recursive formulas
\bsub\label{spin s EOM and delta}
\be
\hspace{-1cm}\dot{Q}_s&=\D_\phi Q_{s-1}-(s+1)(C+u)Q_{s-2},&&\text{for }-1\leq s\leq2,\label{higher spin EOM}\\
\hspace{-1cm}\delta_\xi Q_s&=\left(f\partial_u+Y\partial_\phi+\f{5-s}{2}Y'\right)Q_s+(s+2)f'Q_{s-1},&&\text{for }-2\leq s\leq2.\label{higher spin delta}
\ee
\esub
For $s=-1$, the evolution equation \eqref{higher spin EOM} follows tautologically from the definitions \eqref{definitions curly N and J}, where one can also note that $2\J=\D_\phi\hat{N}$ if $\hat{N}$ is assigned a weight $s=-2$. For $0\leq s\leq 2$, these equations are a rewriting of the evolution equations \eqref{pre-flux-balance} for the mass and the angular momentum, and of the evolution equation for $\Phi_3$ which we had not written out so far. In particular, the evolution equations for $Q_0$ and $Q_1$, or equivalently $\M$ and $\P$, play an important role as they allow to check quite straightforwardly the conservation of the Wald--Zoupas charges \eqref{scalar WZ charge with Y} in the non-radiative regime where $Q_{-2}=0=Q_{-1}$. The equations \eqref{higher spin delta} are a rewriting of the transformation laws~\eqref{covariant transformation laws}, and also provide the transformation law associated with $\Phi_3$. Note that this whole structure is also non-trivial already when allowing only for supertranslations, i.e. when $\psi=0$ and~$Y=Y_0$ as discussed in subsection \ref{subsec: scalar_stranslations}.

Interestingly, one should note that this result about the rewriting of the evolution equations and the transformation laws does not seem to hold in asymptotically-flat spacetimes with a scalar field. This can be seen for example from the fact that in this case $M$ and $P$ have the same conformal weight in \eqref{scalar flat transformation laws}, and therefore cannot be used to build charges $Q_0$ and $Q_1$ transforming as in \eqref{higher spin delta}. The question of whether alternative definitions of the charges $Q_s$ can be given in three-dimensional asymptotically-flat spacetimes with matter is kept for future work.

\subsubsection{Four-dimensional analogy}

Equations \eqref{spin s EOM and delta} are interesting not only for the compact and elegant rewriting of the evolution equations and the transformation laws which they provide, but also for their exact analogy with equations derived in four-dimensional asymptotically-flat spacetimes \cite{Freidel:2021ytz,Geiller:2024bgf}. There, as explained above in subsection \ref{sec:Cotton}, the covariant aspects are identified from the Weyl scalars. They can also be naturally assigned a spin (or helicity) weight in the Newman--Penrose formalism, inherited from the contraction of the tensor indices $a,b,\dots$ on the asymptotic sphere with a complex dyad $m^a$ and its conjugate $\bar{m}^a$. Contractions with these dyads are respectively raising and lowering the spin weight. In this four-dimensional (4d) setup, the spin label of the charges $Q_s$ is therefore traced back to the fact that
\be\label{4d spin s charges}
Q_{-2}^{\text{4d}}=\N_{ab}\bar{m}^a\bar{m}^b,
\quad\,
Q_{-1}^{\text{4d}}=\J_a\bar{m}^a,
\quad\,
Q_0^{\text{4d}}=\M+i\widetilde{\M},
\quad\,
Q_1^{\text{4d}}=\P_am^a,
\quad\,
Q_2^{\text{4d}}=\E_{ab}^1m^am^b.
\ee
Up to the appearance of $\widetilde{\M}$, which does not seem to admit a direct analogue in the present three-dimensional setting, these quantities are manifestly analogous to the charges~\eqref{spin s charges} introduced above, with the role of the spin-2 charge $Q_2^{\text{4d}}$ now played by~\eqref{3d spin-2 charge}. To clarify this correspondence, let us recall that four-dimensional asymptotically flat spacetimes possess an angular metric of the form
\be\label{4d angular metric}
g_{ab}=r^2q_{ab}+rC_{ab}+D_{ab}+\sum_{n=1}^\infty\f{E^n_{ab}}{r^n},
\ee
where $q_{ab}$ is a (not necessarily round) sphere metric, $C_{ab}$ is the gravitational shear, $D_{ab}$ has a trace part determined by the shear and a trace-free part related to logarithmic terms (which is usually set to zero by hand, see footnote \ref{log footnote} above), and $E^n_{ab}$ is the so-called incoming radiation ($E^1_{ab}$ differs from $\E^1_{ab}$ in \eqref{4d spin s charges} by terms cubic in the shear \cite{Freidel:2021ytz,Geiller:2024bgf}). Here, for three-dimensional asymptotically-FLRW spacetimes sourced by a scalar field, the analogue of $g_{ab}$ is $\Phi$. Indeed, the expansion \eqref{fall-offs Phi with Psi0} is of the form
\be\label{3d expansion}
\Phi=2\ln(r+u)+\psi+\f{C}{r}+\f{\Phi_2}{r^2}+\sum_{n=3}^\infty\f{\Phi_n}{r^n}.
\ee
If $q_{ab}$ is a round sphere as in the original construction of \cite{Bondi:1960jsa,Bondi:1962px}, the asymptotic symmetries contain no superrotations. This is the analogue of what happens when setting $\psi$ to zero in the present context. At subleading order, $C_{ab}$ is clearly the analogue of $C$. Then, $D_{ab}$ is also analogous to $\Phi_2$ in the sense that these two fields contain generically a contribution determined by the shear and an additional contribution which, when non-vanishing, leads to logarithmic terms in the solution space. Finally, the role of the incoming radiation  $E^n_{ab}$ (which can be related to the multipole moments of the source \cite{Compere:2022zdz}) is played here by $\Phi_{n\geq3}$, and we have indeed that $E^1_{ab}$ and $\Phi_3$ can be used to define spin-2 charges $Q_2$.

In four-dimensional asymptotically-flat spacetimes, the role of $\D_\phi$ in \eqref{higher spin EOM} played by the Newman--Penrose ``edth'' operator $\eth$, whose action on quantities of spin $s$ is given in terms of the angular dyad by $\eth=m^a\partial_a-sD_am^a$, where $D_a$ is the covariant derivative with respect to $q_{ab}$. One can recognize once again a strong analogy between the divergence terms in $\eth$ and the term $\psi'$ in~\eqref{covariant D phi}, especially since the freedom in $\psi$ and $q_{ab}$ is related to the presence of superrotations.

In summary, we have shown that although our three-dimensional solution space is constructed solely in terms of scalar quantities, these can be mapped to ``spin-$s$'' charges \eqref{spin s charges} to obtain equations \eqref{spin s EOM and delta} and reveal an interesting analogy with the structure of four-dimensional asymptotically-flat spacetimes (where the spin label $s$ is inherited from non-trivial tensorial structure on the transverse sphere, as in \eqref{4d spin s charges}). This is particularly intriguing because this higher-spin structure and evolution equations similar\footnote{In four-dimensional asymptotically-flat spacetimes, up to normalization conventions, this equation takes the form $\dot{Q}_s=\eth Q_{s-1}-(s+1)CQ_{s-2}$ where $C\coloneqq C_{ab}m^am^b$. There is therefore no shift in $u$ of the term involving $C$, which here can be traced back to the fact that we are in an FLRW background.} to \eqref{higher spin EOM} have recently been used to study the $w_{1+\infty}$ symmetry algebra of four-dimensional self-dual gravity \cite{Ball:2021tmb,Strominger:2021lvk,Adamo:2021lrv,Freidel:2021ytz,Compere:2022zdz,Geiller:2024bgf,Kmec:2024nmu,Cresto:2024fhd,Cresto:2024mne}. This begs the question of whether a similar symmetry algebra can be discovered in (a subsector of) the radiative solution space studied here, and also in the case of three-dimensional asymptotically-flat spacetimes coupled to a scalar or Maxwell field. One can also wonder whether four-dimensional asymptotically-FLRW spacetimes can exhibit such a symmetry algebra. We leave these interesting investigations for future work.

\subsubsection{Newman--Penrose charges}

In order to complete the analogy with the four-dimensional asymptotically-flat case, let us now extend the pattern \eqref{higher spin EOM} to $s=3$. In the four-dimensional case, this result follows directly from the Newman--Penrose Bianchi identity encoding the time evolution of $\Psi_0^1$, which is the first subleading term in the expansion of the Weyl scalar
\be
\Psi_0^\text{4d}=\f{\Psi_0^0}{r^5}+\f{\Psi_0^1}{r^6}+\O(r^{-7}).
\ee
It was shown in \cite{Newman:1965ik,Newman:1968uj} that $\Psi_0^1$ satisfies
\be\label{4d Psi01 dot}
\dot{\Psi}_0^1=-\bar{\eth}\big(\eth Q_2^\text{4d}-4CQ_1^\text{4d}\big),
\ee
where $Q_2^\text{4d}=\Psi_0^0$ and $Q_1^\text{4d}=\Psi_1^0$. The fact that this evolution equation gives a total $\bar{\eth}$ derivative enables to show that $\Psi_0^1$, when smeared with a proper spin-weighted spherical harmonic, leads to 10 exactly conserved quantities \cite{Newman:1965ik,Newman:1968uj}. These are the so-called non-linear Newman--Penrose charges. Following \cite{Freidel:2021ytz,Geiller:2024bgf}, one can also define a spin-3 charge via $\Psi_0^1\coloneqq-\bar{\eth}Q_3^\text{4d}$ and find that this quantity follows the same recursive evolution equation as for $-1\leq s\leq2$.

In the four-dimensional case, $\Psi_1^0$ is related to the subleading term $E^2_{ab}$ in \eqref{4d angular metric}. Relying on the analogy with expansion \eqref{3d expansion} developed above, we can then expect that the three-dimensional version of \eqref{4d Psi01 dot} will involve $\Phi_4$. Remarkably, this turns out to be the case, and one finds that the quantity
\be\label{NP charge}
F_4\coloneqq-48\Phi_4-72u\Phi_3+12u^3C
\ee
satisfies the evolution equation
\be
\dot{F}_4=\big(\partial_\phi+\psi'\big)\big(\D_\phi Q_2-4(C+u)Q_1\big).
\ee
Defining a spin-3 charge as $F_4\coloneqq\big(\partial_\phi+\psi'\big)Q_3$ extends the pattern~\eqref{higher spin EOM} to the case $s=3$. When $\psi=0$, it follows immediately that the integral of $F_4$ over the celestial circle is exactly conserved, since its time evolution reduces to the integral of a total angular derivative and therefore vanishes. This quantity can thus be identified with the three-dimensional Newman--Penrose charge for asymptotically FLRW spacetimes coupled to a scalar field.

An interesting question which we have not addressed here is that of the geometrical origin of the evolution equations \eqref{higher spin EOM}. In the four-dimensional case, these equations appear in the Newman--Penrose formalism as the Bianchi identities on the Weyl scalars. In the three-dimensional Newman--Penrose formalism however \cite{Hall:1987vw,Aliev:1995cf,Milson:2012ry,Barnich:2016lyg}, the Bianchi identities do not provide non-trivial information. Instead of being formulated in terms of the Weyl tensor (which is vanishing in the three-dimensional case), they just provide consistency conditions on the derivatives of the Ricci scalars. In the notations of e.g. \cite{Barnich:2016lyg}, the evolution equation for the mass aspect comes instead from the spin coefficient equation on $\mu$ (see equation (A.7) there). The question is therefore how to rearrange the three-dimensional Newman--Penrose equations in order to access the evolution equations \eqref{higher spin EOM} built from the Cotton scalars. We leave this investigation for future work.

%%%%%%%%%%%%%%%%%%%%%%%%%%%%%%%%%%%%%%%%%%%%%%%%%%%%%%%%%%%%%%%%%%%

\section{Coupling to a Maxwell field}
\label{sec:Maxwell}

In the previous section we have studied asymptotically-FLRW spacetimes sourced by a scalar field. It is now natural to consider Einstein--Maxwell theory, not so much for the increase in complexity, which as we shall see is minor, but because Maxwell and scalar fields are dual in three dimensions and encode the same dynamics. In particular, it is interesting to ask how this duality is implemented at the level of the solution space and of the charges.

\subsection{Generalities}

In three spacetime dimensions, every anti-symmetric tensor $F_{\mu\nu}=F_{[\mu\nu]}$ can be written in the form~$F_{\mu\nu}=\eps_{\mu\nu\rho}\tilde{F}^\rho$. Maxwell's equations then imply that $\tilde{F}$ is the gradient of some scalar field $\Phi$. Indeed, we have
\be
\nabla^\mu F_{\mu\nu}=\eps_{\mu\nu\rho}\nabla^\mu\tilde{F}^\rho=0\qquad\Leftrightarrow\qquad\nabla_{[\mu}\tilde{F}_{\rho]}=0\qquad\Leftrightarrow\qquad\tilde{F}_\mu\propto\nabla_\mu\Phi.
\ee
The Maxwell Bianchi identities are then equivalent to the equation of motion of the free scalar field, as
\be
\nabla_{[\mu}F_{\nu\rho]}=0\qquad\Leftrightarrow\qquad\nabla_\mu\tilde{F}^\mu=0\qquad\Leftrightarrow\qquad\Box\Phi=\nabla_\mu\nabla^\mu\Phi=0.
\ee
Given the Maxwell field strength $F_{\mu\nu}=\partial_\mu A_\nu-\partial_\nu A_\mu$, here we are going to use the duality map
\be\label{dictionary}
F_{\mu\nu}=\f{1}{2}\eps_{\mu\nu\rho}\partial^\rho\Phi=\f{1}{2}\sqrt{-g}\,\epsilon_{\mu\nu\rho}\partial^\rho\Phi
\ee
in order to import the previous results about $\Phi$ and discuss the asymptotic structure of FLRW spacetimes coupled to the spin-1 gauge field $A_\mu$. Under this duality, the stress-energy tensor \eqref{scalar stress tensor} of the scalar theory becomes
\be
T_{\mu\nu}[\Phi,g]=\f{1}{2}\nabla_\mu\Phi\nabla_\nu\Phi-\f{1}{4}g_{\mu\nu}\nabla^\alpha\Phi\nabla_\alpha\Phi
\stackrel{\eqref{dictionary}}{=}
2F_{\mu\alpha}F_\nu{}^\alpha-\f{1}{2}g_{\mu\nu}F_{\alpha\beta}F^{\alpha\beta}=T_{\mu\nu}[A,g],
\ee
and is therefore equivalent to the stress-energy tensor derived from the Maxwell Lagrangian\footnote{Note that one should \textit{not} use the duality map \eqref{dictionary} naively at the level of the Lagrangians, otherwise this produces a mismatch of sign and sends $L_\text{Maxwell}\to-L_\text{scalar}$. This is due to the fact that the duality holds on-shell.}
\be\label{Maxwell Lagrangian}
L_\text{Maxwell}=-\f{1}{2}\sqrt{-g}\,F_{\mu\nu}F^{\mu\nu}.
\ee
This implies that three-dimensional Maxwell theory can also be viewed as a perfect fluid with~$k=1$. The metric satisfies the same Einstein field equations sourced by matter as described in subsection~\ref{sec:Bondi hierarchy}.

In order to compute the asymptotic charges associated with the infinitesimal transformations $\delta_{(\xi,\epsilon)}A_\mu=\pounds_\xi A_\mu+\partial_\mu\epsilon$, we need the equivalent of formula \eqref{IW general Einstein scalar} for Einstein--Maxwell theory \cite{Bosma:2023sxn}. When coupling the Einstein--Hilbert Lagrangian to the Maxwell Lagrangian \eqref{Maxwell Lagrangian}, the total presymplectic potential becomes
\be
\theta^\mu=\theta^\mu_\text{EH}+\theta^\mu_\text{Maxwell}=\f{1}{2}\sqrt{-g}\,\Big(g^{\alpha\beta}\delta\Gamma^\mu_{\alpha\beta}-g^{\mu\alpha}\delta\Gamma^\beta_{\alpha\beta}-4F^{\mu\nu}\delta A_\nu\Big).
\ee
The Komar--Maxwell charge is then given by the gravitational contribution \eqref{IW general Einstein scalar} plus the Maxwell contribution, adding up to
\be
K^{\mu\nu}_{(\xi,\epsilon)}=-\f{1}{2}\sqrt{-g}\,\Big(2\nabla^{[\mu}\xi^{\nu]}+4F^{\mu\nu}\big(\xi^\alpha A_\alpha+\epsilon\big)\Big).
\ee
Finally, this Noether charge can be combined with the symplectic potential to form the Iyer--Wald charge aspect
\be\label{Maxwell IW charge}
\slashed{\delta}k^{\mu\nu}_{(\xi,\epsilon)}=\delta K^{\mu\nu}_{(\xi,\epsilon)}-K^{\mu\nu}_{(\delta\xi,\delta\epsilon)}+2\xi^{[\mu}\theta^{\nu]}.
\ee
Importantly, the second term in this expression subtracts contributions coming from the possible field-dependence of the gauge parameters $\xi$ and $\epsilon$. This is such that, in the combination of the first two terms, the variation $\delta$ does not act on the parameters, as required from the fact that the Iyer--Wald charge is the contraction of infinitesimal gauge transformations with the symplectic structure, and therefore does not contain variations of the gauge parameters (even when they are field-dependent).

It should be noted that although the scalar and Maxwell theories are dynamically equivalent in three-dimensional spacetimes, Maxwell theory remains a gauge theory while the scalar field exhibits no gauge freedom. In four spacetime dimensions, the duality is instead between a scalar field and a 2-form gauge theory, and was studied in order to provide a symmetry interpretation of the scalar soft theorems \cite{Campiglia:2017dpg,Francia:2018jtb,Campiglia_2019,Henneaux_2019,Geiller:2021gdk}. Finally, it is interesting to point out that under the duality~\eqref{dictionary} the presymplectic potentials $\theta_\text{Maxwell}$ and $\theta_\text{scalar}$ are not equal. It is straightforward to check this with the on-shell solution space below. This well-known subtlety is common to situations where different Lagrangians describe the same equations of motion using different variables. A typical example of this is in the case of gravity with metric and tetrad variables, described respectively by the Einstein--Hilbert and Einstein--Cartan Lagrangians, whose symplectic potentials differ by a corner contribution \cite{DePaoli:2018erh,Oliveri:2019gvm,Freidel:2020svx,Freidel:2020xyx}. This difference in the symplectic potentials will result in different expressions for the charges of Maxwell theory.

\subsection{Solution space}

In order to build the solution space describing asymptotically-FLRW metrics \eqref{metric ansatz} coupled to a Maxwell field, we are going to work in the radial gauge and set
\be
A_r=0.
\ee
First of all, one can note that the background solution representing the Maxwell analogue of the scalar result \eqref{FLRW-scalar backg Bondi coord.} is
\be
\beta=0,
\q
V=-r,
\q
U=0,
\q
A_\mu\,\de x^\mu=-\f{r^2}{2}\de\phi.
\ee
Our goal is to describe perturbations on top of this background solution, subject to the boundary conditions of section~\ref{sec:asymptotically-FLRW} and allowing for non-trivial superrotations.

By exploiting the duality between the Maxwell and scalar fields, one can write down the solution space for Einstein--Maxwell theory without solving explicitly the corresponding field equations. Instead, one can simply import the results of the previous section on Einstein-scalar and integrate~\eqref{dictionary} to obtain the gauge fields. This leads to
\bsub\label{A from Psi}
\be
&F_{ru}=\partial_rA_u=\f{1}{2}\sqrt{-g}\,\epsilon_{ru\phi}\partial^\phi\Phi&&\qquad\Rightarrow\qquad&&A_u=-\f{1}{2}\int\de r\sqrt{-g}\,\partial^\phi\Phi+ A_u^0,\\
&F_{r\phi}=\partial_rA_\phi=\f{1}{2}\sqrt{-g}\,\epsilon_{r\phi u}\partial^u\Phi&&\qquad\Rightarrow\qquad&&A_\phi=\f{1}{2}\int\de r\sqrt{-g}\,\partial^u\Phi+A_\phi^0,\\
&F_{\phi u}=\f{1}{2}\sqrt{-g}\,\epsilon_{\phi ur}\partial^r\Phi&&\qquad\Rightarrow\qquad&&\partial_uA_\phi=\partial_\phi A_u-\f{1}{2}\sqrt{-g}\,\partial^r \Phi,
\ee
\esub
where $A_\phi^0=A_\phi^0(u,\phi)$ and $A_u^0=A_u^0(u,\phi)$ are radial integration constants that encode the residual gauge freedom of the Maxwell field. Importantly, the relation \eqref{dictionary} shows that the field $\psi(\phi)$ does not enter the Einstein--Maxwell dynamics directly, but appears instead only with an angular derivative. This motivates the introduction of the notation
\be
\psi'\eqqcolon E(\phi),
\ee
where $E$ will play later on the role of electric charge aspect. By integrating explicitly \eqref{A from Psi} with the scalar field $\Phi$ given in subsection \ref{subsec. superrot-solution space}, we obtain
\bsub\label{Maxwell solution}
\be
A_u&=-\f{3}{2}Er+A_u^0+\f{1}{r}\left(\f{1}{4}C'(u+3C)+\f{3}{16}EC(2u+C)-P\right)+\O(r^{-2}),\\
A_\phi&=-\f{1}{2}r^2+\f{1}{2}Cr+A_\phi^0+\O(r^{-1}).
\ee
\esub
It is interesting to compare this result with the radiative Maxwell field in asymptotically-flat spacetimes \cite{Bosma:2023sxn}. There, the Coulombic term representing the electric charge aspect appears in a logarithmic $\ln(r)$ term, while the radiative term representing the shear appears at order $\sqrt{r}$. By contrast, here the Coulombic and radiative fields $E$ and $C$ appear both at order $r$. This is a consequence of the presence of a scale factor $a(u,r)$ in the metric, which in turn affects the radial order of the radiative branches of the solution space. Furthermore, the gauge field contains no logarithmic terms since we have imposed a condition removing such terms in the case of the scalar field. Finally, the gauge field contains a pure gauge contribution encoded in $A_u^0$ (that was denoted $G$ in~\cite{Bosma:2023sxn}). The time dependence of this term is left completely unconstrained. The terms appearing in the expansion of $A_\phi$ have their time evolution determined by the Maxwell equation $\nabla^\mu F_{\mu\phi}=0$, starting with
\be
\dot{A}_\phi^0=2M+\big(A_u^0\big)'-\f{1}{4}\Big(EC'+uE^2-uN+C\big(E^2+3E'-1\big)\Big),
\ee
where we recall that here $N=\dot{C}$. The subleading terms in $A_\phi$ depend on $\Phi_{n\geq3}$ and have a time evolution set by $\nabla^\mu F_{\mu\phi}=0$, which is furthermore equivalent to the equations on $\dot{\Phi}_n$ imposed by the Klein--Gordon equation.

\subsection{Residual symmetries}

When turning to the asymptotic symmetries, Maxwell theory is clearly different from the scalar field since it exhibits a gauge symmetry $A_\mu\mapsto A_\mu+\partial_\mu\epsilon$. By contrast, the scalar is only invariant under global shifts. Since we are working in radial gauge $A_r=0$, we must ensure that residual gauge transformations $\delta_{(\xi,\epsilon)}A_\mu=\pounds_\xi A_\mu+\partial_\mu\epsilon$ also preserve this gauge choice. This can be achieved if the U(1) gauge parameter satisfies $\pounds_\xi A_r+\partial_r\epsilon=0$ with $\xi$ given by \eqref{AKV components}. This leads to the expression
\be\label{epsilon U(1)}
\epsilon(u,r,\phi)
&=\alpha(u,\phi)+f'\int_r^\infty\f{e^{2\beta}A_\phi}{\tilde{r}^2}\de\tilde{r}\cr
&=\f{r}{2}f'+\alpha(u,\phi)+\f{1}{16r}\big(16 A_\phi^0+6uC+5C^2\big)f'+\O(r^{-2}),
\ee
where $\alpha(u,\phi)$ parametrizes the remaining U(1) gauge freedom of the Maxwell field. In other words, $\alpha$ is the symmetry parameter for the asymptotic ``superphase rotations''.

Taking into account the new U(1) parameter, we find that the bracket of the symmetry transformations $\delta_{(\xi,\epsilon)}$ is $\big[\delta_{(\xi_1,\epsilon_1)},\delta_{(\xi_2,\epsilon_2)}\big]=-\delta_{(\xi_{12},\epsilon_{12})}$ with
\be
\xi_{12}\coloneqq\big[\xi_1,\xi_2\big]_*=\big[\xi_1,\xi_2\big]-\delta_{\xi_1}\xi_2+\delta_{\xi_2}\xi_1,
\q\q
\epsilon_{12}\coloneqq\pounds_{\xi_1}\epsilon_2-\delta_{(\xi_1,\epsilon_1)}\epsilon_2-(1\leftrightarrow2).
\ee
The symmetry parameters on $\mathscr{I}^+$ form a Maxwell extension of the $\mathfrak{bms}_3^k$ algebra with $k=1$. We have
\be
\big[(f_1,Y_1,\alpha_1),(f_2,Y_2,\alpha_2)\big]=(f_{12},Y_{12},\alpha_{12}),
\ee
with $f_{12}$ and $Y_{12}$ given in \eqref{eq. bms3like} for $k=1$ and
\be
\alpha_{12}=f_1\dot{\alpha}_2+Y_1\alpha_2'-(1\leftrightarrow2).
\ee
We note that the transformation of the Maxwell gauge parameter is unaffected by the cosmological deformation due to the parameter $k$ when compared to the flat case \cite{Barnich:2015jua,Bosma:2023sxn}, at the difference with the translations.

The action of the residual symmetries parametrized by the vector field \eqref{bmsk=1 diffeos} and the U(1) parameter \eqref{epsilon U(1)} on the Maxwell field at leading order is given by
\bsub
\be
\delta_{(\xi,\epsilon)}C&=\left(f\partial_u+Y\partial_\phi+\f{1}{2}Y'\right)C+f+f''-2f'E+\f{1}{2}uY',\\
\delta_{(\xi,\epsilon)}N&=\left(f\partial_u+Y\partial_\phi+2Y'\right)N+2Y'-3Y''E+\f{3}{2}Y''',\\
\delta_{(\xi,\epsilon)}E&=(YE)'-Y'',\label{transformation E}\\
\delta_{(\xi,\epsilon)}A_u^0
&=\left(f\partial_u+Y\partial_\phi+\f{3}{2}Y'\right)A_u^0+\f{3}{4}\left(f+\f{1}{2}uY'-f''\right)E+\f{3}{2}f'E'\cr
&\pe+\f{3}{4}f'E^2-\f{3}{8}Y''C+\f{1}{4} f'N+\dot{\alpha},\label{delta Au0}\\
\delta_{(\xi,\epsilon)}A_\phi^0&=\big(f\partial_u+Y\partial_\phi+Y'\big)A_\phi^0-\f{1}{4}f(u+C)-\f{1}{8}\big(uY'+2f''\big)C-\f{3}{4}f'C'\cr
&\pe-\f{1}{4}f'CE+f'A_u^0-\f{1}{4}uf'-\f{1}{8}u^2Y'+\alpha',
\ee
\esub
where we use the notation $\delta_{(\xi,\epsilon)}$ to emphasize the presence of an additional symmetry parameter when compared to the scalar case. One can see that the transformation laws for $C$ and $N$ are the same as in the scalar case \eqref{scalar transformations with Y} with the identification $\psi'=E$. When using this in \eqref{transformation psi}, we recover the transformation law \eqref{transformation E} for $E$. Since the metric does not transform under the U(1) symmetry, the action of $\delta_{(\xi,\epsilon)}$ on $M$ and $P$ is still given by \eqref{deltaM} and \eqref{deltaP} as in the case of scalar matter.

In the analysis of the charges performed below, we will see that $E$ appears as the electric charge aspect associated with the residual U(1) symmetry parameter $\alpha$. Interestingly, the transformation law \eqref{transformation E} therefore shows that a non-vanishing electric charge aspect is required in order to obtain unconstrained superrotations $Y(\phi)$. As in the scalar case, the presence of these superrotations implies that $N\coloneqq\dot{C}$ does not transform homogeneously, and a notion of covariant news can be identified by studying the vacuum structure of Einstein--Maxwell theory.

\subsection{Vacuum structure and covariant news}

We now discuss, as a generalization of the construction presented for the scalar field, the structure of the vacuum for Einstein--Maxwell theory. This is obtained once again by acting with a finite diffeomorphism which brings a background solution to Bondi gauge. In conformal coordinates $\tilde{x}^\mu=(\tau,\rho,\theta)$, we start from the FLRW background coupled to a Maxwell field \eqref{FLRW-scalar backg Bondi coord.}, which reads
\bsub\label{starting_g and A}
\be
\de s^2&=\tilde{g}_{\mu\nu}\de\tilde{x}^\mu\de\tilde{x}^\nu=a(\tau)^2\big(\!-\!\de\tau^2+\de\rho^2+\rho^2\de\theta^2\big),\qquad a(\tau)=\tau,\\
\tilde{A}_\mu\de\tilde{x}^\mu&=-\f{\rho^2}{2}\de\theta.
\ee
\esub
As explained in appendix \ref{Appendix vacuum}, we then map this configuration to Bondi gauge by means of the finite diffeomorphism $\tilde{x}^\mu=(\tau,\rho,\theta)\mapsto x^\mu=(u,r,\phi)$ given in \eqref{sol finite diffeo}.

An additional subtlety appears in the case of a Maxwell field. Indeed, under the action of the diffeomorphism the gauge potential transforms as
\be
\tilde{A}_\mu(\tilde{x})\mapsto A_\mu(x)=\f{\partial\tilde{x}^\alpha}{\partial x^\mu}\tilde{A}_\alpha\big(\tilde{x}(x)\big),
\ee
which generically spoils the radial gauge condition $A_r=0$. The radial gauge can then be restored thanks to a compensating gauge transformation $\delta_\lambda A_\mu=\partial_\mu\lambda$, chosen such that the final configuration lies in Bondi gauge for the metric and in the radial gauge for the Maxwell field. The required gauge parameter takes the form
\be
\lambda=\f{1}{4}r\big(2\T'+3(u+\T)\W'\big)+\A(u,\phi)+\O(r^{-1}),
\ee
where $\A(u,\phi)$ is an arbitrary function. This new superphase rotation field $\A(u,\phi)$ is characteristic of the Maxwell case, and it encodes the residual U(1) gauge symmetry. By contrast with the scalar case, the vacuum orbit therefore acquires an additional functional degree of freedom associated with this U(1) gauge symmetry. The vacuum structure is thus given by a three-parameter family of configurations labeled by $\cT(\phi)$, $\cW(\phi)$ and $\A(u,\phi)$. At leading order, studying the expression for the components \eqref{Maxwell solution} in the vacuum orbits reveals that
\bsub\label{Evac}
\be
E&\stackrel{\text{vac}}{=}E_\text{vac}\coloneqq-\W',\\
\label{Cvac Maxwell}
C&\stackrel{\text{vac}}{=}C_\text{vac}\coloneqq2\T'\W'+\T''+\big(\T+u\big)\left(e^{2\W}+\f{3}{4}\left[\big(\W'\big)^2+2\W''\right]\right)-u,\\
\label{A0u vac}
A^0_u&\stackrel{\text{vac}}{=}\dot{\A}-\f{1}{16}\Big(6\T\W'+\T'\Big[4-4e^{2 \cW}+9\big(\W'\big)^2+6\W''\Big]-6\W'\T''\Big),\\
A_\phi^0 &\stackrel{\text{vac}}{=}\A'+a_\phi^0[\T,\W],
\ee
\esub
where the explicit expression for $a_\phi^0$ is not necessary for what follows. Equation~\eqref{A0u vac} shows that $A_u^0$ is pure gauge along the vacuum orbit. In particular, it can always be set to zero by an appropriate choice of the superphase rotation parameter $\A(u,\phi)$. Finally, since the metric is invariant under the U(1) gauge transformations, the metric on the vacuum orbit is the same as in the scalar case, and it is given again by \eqref{vacuum-valued bondi param}.

The upshot of this calculation of the vacuum structure is that the vacuum shear $C_{\text{vac}}$ in the Maxwell case \eqref{Cvac Maxwell} is the same as \eqref{Cvac} for the scalar field. This justifies a posteriori the use of the duality \eqref{dictionary}, and also indicates that the covariant news tensor in the case of Maxwell matter will have a similar structure as in the scalar case. Following the same strategy as in the scalar analysis, we would like to define a covariant news tensor that vanishes in the vacuum. However, a subtlety of the Maxwell case is that not all the quantities entering $C_\text{vac}$ can be expressed in terms of the fields in the solution space. To understand this obstruction, one should recall that although we have the identification $E\coloneqq\psi'$, there is no analogue of $\psi$ itself in the Maxwell solution space. As a consequence, $\W$ cannot be identified with a field in the solution space, and has to be thought of as a superboost field introduced by hand in order to achieve covariance of the news. We then consider the definition
\be\label{News Maxwell}
\hat{N}
&\coloneqq\dot{C}-\dot{C}_\text{vac}\big|_{\W'=-E}\cr
&\,=\dot{C}+\f{3}{2}E'-\f{3}{4}E^2+1-e^{2\W}.
\ee
Using the transformation of $\W$ given in \eqref{Transfo W}, this news tensor transforms homogeneously as
\be\label{transfo Maxwell News}
\delta_{(\xi,\epsilon)}\hat{N}=\big(f\partial_u+Y\partial_\phi+2Y'\big)\hat{N}.
\ee
In conclusion, we arrive at a similar covariant news as in the scalar case, although here $\W$ has to appear explicitly since $\psi$ does not exist in the Einstein--Maxwell solution space.

\subsection{Charge algebra}

In the case of Maxwell matter, the asymptotic charges must now be computed with the $(ur)$ component of the formula \eqref{Maxwell IW charge}, where the symmetry generators are the vector field $\xi$ and the U(1) parameter $\epsilon$. The functions which parametrize these symmetry generators are $f$, $Y$, and $\alpha$.

With an arbitrary split between integrable and non-integrable parts, the result is finite and given by \eqref{bare split IW Maxwell 1} and \eqref{bare split IW Maxwell 2}. This charge can be further split between a purely gravitational part where the symmetry parameters are $f$ and $Y$, and a U(1) part with symmetry parameter $\alpha$. For the former, an integrable part can be chosen as in the scalar case following the construction presented in subsection~\ref{sec:charge algebra with Y}. The U(1) part on the other hand is given by
\be
Q_\alpha=\oint\alpha E.
\ee
As expected, for $\alpha=1$ this measures the total electric charge. For a generic value, a peculiar feature of this contribution is that $\alpha=\alpha(u,\phi)$ has an arbitrary dependence on retarded time. Therefore, although $E$ is time-independent, we get that
\be\label{Q alpha dot}
\dot{Q}_\alpha=\oint\dot{\alpha}E,
\ee
and this contribution cannot be vanishing even in the absence of radiation. The transformation law \eqref{delta Au0} shows that the arbitrary time dependence in $\alpha$ is related to the presence of $A_u^0$ in the solution space. Indeed, since this transformation law is not homogeneous, setting $A_u^0=0$ would determine the form of $\dot{\alpha}$. However, even in this case \eqref{Q alpha dot} is non-vanishing. This suggests that the Wald--Zoupas charges which are conserved in the absence of radiation should be defined only with the purely gravitational contribution parametrized by $f$ and $Y$, or in other words with $\alpha=0$ for the U(1) part. Note that in light of the duality between Maxwell and the scalar, since there the electric charge becomes $E=\psi'$, we get that $Q_{(\alpha=1)}=0$.

Following the construction presented above in the case of scalar matter, one can select an integrable part in the split \eqref{bare split IW Maxwell 1} and \eqref{bare split IW Maxwell 2} by implementing criteria 1) and 2) of section \ref{sec:charge algebra with Y}. A natural candidate is then given by the analogue of \eqref{scalar WZ charge with Y}, which here is
\be
Q_\xi=\f{1}{2}\oint\Big(4T\M-3Y\P+2uY\big(2\M E-\M'\big)\Big).
\ee
This integrable part and the corresponding split can then be used to compute the Barnich--Troessaert bracket, which closes up to a field-dependent cocycle given in \eqref{Coc Einstein-Maxwell}. The study of an improved prescription for the charges, leading potentially to a field-independent cocycle, is left for future work, and should first be understood in the case of scalar matter.
 
%%%%%%%%%%%%%%%%%%%%%%%%%%%%%%%%%%%%%%%%%%%%%%%%%%%%%%%%%%%%%%%%%%%

\section{Conclusion}
\label{sec:conclusion}

In this work we have taken a new step towards the understanding of asymptotically-cosmological spacetimes in general relativity. For this, we have studied detailed properties of three-dimensional spatially flat and decelerating asymptotically-FLRW spacetimes in the presence of scalar and Maxwell matter. This provides a simplified arena where conceptual and technical issues arising in the physically-relevant four-dimensional case can already be investigated. For this analysis, we have first exhibited gauge-fixing and fall-off conditions which lead to a $\mathfrak{bms}_3^k$ algebra of asymptotic symmetries. This is the three-dimensional counterpart of the BMS-like symmetries found in the four-dimensional analysis of \cite{Bonga:2020fhx,Enriquez-Rojo:2021blc}. In order to study explicit realizations of this algebra we have then focused on the coupling to a massless scalar field, which corresponds to the case $k=1$. This particular example enables to understand various subtleties. First, the explicit analysis and resolution of the non-vacuum Einstein field equations with the ansatz \eqref{metric ansatz} reveals that the Coulombic data, or the Bondi mass and angular momentum aspects, enter the line element at a radial order which is different from the ansatz considered in \cite{Enriquez-Rojo:2022onp,Enriquez-Rojo:2020miw,Enriquez-Rojo:2021blc}. Indeed, in these references the line element is also of the form $\de s^2=a^2\de\bar{s}^2$, but the mass and angular momentum have been introduced in~$\de\bar{s}^2$ at the same radial order as for asymptotically-flat spacetimes. Our results indicate that this is incorrect in the three-dimensional case, and upcoming work will also revisit the construction of the four-dimensional solution space. This is the first message of the present work, namely that a careful study of the solution space in the presence of a specific matter sector reveals subtleties in the identification of the Coulombic data as radial integration constants. For a generic $k$, the line element must actually be expanded in both $r$ and $a$ in order to access the radial integration constants, and this is clearly at odds with the ansatz considered in \cite{Enriquez-Rojo:2022onp,Enriquez-Rojo:2020miw,Enriquez-Rojo:2021blc}.

As a second lesson from this work, we have seen that an extra mode in the radial expansion of the radiative scalar field is required in order to have access to non-trivial superrotations. When mapping the scalar field to its dual Maxwell field, this mode is related via its angular derivative to the electric charge aspect. When only supertranslations are present, i.e.\ when this extra mode is turned off as in \eqref{fall-offs Phi no Psi0}, one can define Wald--Zoupas charges and show that their Barnich--Troessaert bracket closes without central extension. When superrotations are allowed, the presence of the extra mode of the scalar field in \eqref{fall-offs Phi with Psi0} leads to intricacies in the solution space and the charges. This is similar to the complications which appear when considering an arbitrary leading sphere metric $q_{ab}$ in four-dimensional asymptotically-flat spacetimes, which is the setup of the generalized BMS group. In particular, the presence of superrotations spoils the homogeneous transformation properties of the naive news $N=\dot{C}$. In order to remedy this, we have computed along the lines of~\cite{Compere:2016jwb,Compere:2018ylh} the action of finite diffeomorphisms leading to the Bondi gauge when acting on the vacuum FLRW spacetime in conformal coordinates. This produces a line element in Bondi gauge including a supertranslation and a superboost field, and the latter can be used to construct a covariant news~$\hat{N}$ given in~\eqref{News}. This analysis allowed us to define a covariant mass and angular momentum as the boundary charges associated, respectively, to time-like asymptotic translations and asymptotic rotations. These charges vanish on vacuum configurations and are conserved in the absence of radiation (defined as the vanishing of the covariant news). This is a first important step towards a satisfactory definition of energy for the fluctuations even in the absence of a time-like Killing vector for the non-stationary background, as well as towards an implementation of the Wald--Zoupas prescription in this setup. While our prescription for the charges leads to a field-dependent cocycle, we are hopeful that the model is rich enough in order to accommodate a covariance analysis along the lines of~\cite{Odak:2022ndm,Rignon-Bret:2024gcx,Rignon-Bret:2024wlu}.

Interestingly, the analogy between the present model and four-dimensional asymptotically-flat spacetimes also reveals itself when studying the evolution equations for the covariant mass and angular momentum aspects, and the subleading modes in the scalar field. We have indeed shown in subsection~\ref{sec:subleading} that these evolution equations can partly be written in the compact recursive form~\eqref{higher spin EOM}, which closely resembles the recursion equation that has been used in \cite{Freidel:2021ytz,Geiller:2024bgf} to study the $w_{1+\infty}$ algebra of celestial symmetries in the four-dimensional case. As one could anticipate, here the covariant aspects which satisfy this compact form of the evolution equations appear as components of the Cotton tensor instead of the Weyl tensor, which vanishes identically in three dimensions. These evolution equations also enable to show that subleading terms in the expansion of the scalar field can be used to define exactly conserved Newman--Penrose charges \eqref{NP charge}. To the best of our knowledge this is the first time that an example of such charges is given in three-dimensional gravity.

Having used the present model in order to gather insights into the structure of three- and four-dimensional asymptotically-FLRW spacetimes, there are several interesting directions to explore, on which there is currently work in progress:
\begin{itemize}
%%%%%%%%%%%%%%%%%%%%%%%%
\item\textbf{Four-dimensional asymptotically-FLRW spacetimes.}
The most natural continuation of this work is to revisit the construction of four-dimensional asymptotically-FLRW spacetimes presented in \cite{Enriquez-Rojo:2020miw,Enriquez-Rojo:2021blc,Enriquez-Rojo:2022onp,Enriquez-Rojo:2022ntu}. Indeed, we have argued that the mass and angular momentum aspects which are identified in these references are actually not the quantities appearing as radial integration constants when solving the hypersurface Einstein equations. Our three-dimensional example clearly demonstrates this subtlety. This implies that the whole four-dimensional analysis should be carried out again, with a new derivation of the solution space, the evolution equations, and the asymptotic charges. This will then give the proper framework to study, e.g., the infrared triangle \cite{Creminelli:2024qpu} in asymptotically-FLRW$_4$ spacetimes. For concreteness and in order to access a concrete matter Lagrangian for the computation of the charges, in analogy with the current work these investigations could first be carried out with the perfect fluid stress-energy tensor of a scalar field. For Maxwell fields, one could either study the case of radiation with equation of state parameter $w=1/3$, or go beyond the perfect fluid description and solve the Einstein equations sourced by the four-dimensional Maxwell stress-energy tensor.
%%%%%%%%%%%%%%%%%%%%%%%%
\item\textbf{Memory effects and infrared triangle.}
An interesting question is that of the fate of the infrared triangle in asymptotically-cosmological spacetimes. For asymptotically-flat spacetimes the triangle is a correspondence between asymptotic symmetries, soft theorems, and memory effects \cite{Strominger:2017zoo}. It has been suggested that in the cosmological context this could generalize to a correspondence between adiabatic modes, consistency relations, and cosmological memory respectively \cite{Creminelli:2024qpu}. While some of these aspects have already been studied on their own (see e.g.~\cite{Weinberg:2003sw,Hinterbichler:2012nm,Hinterbichler:2013dpa,Tolish:2016ggo,Pajer:2017hmb,Jokela:2022rhk,Jana:2023kyq,Jokela:2023suz}), a full clear picture as in the asymptotically-flat case is still missing. At least for the study of memory and asymptotic symmetries, an analysis in the ``cosmological Bondi gauge'' \eqref{metric ansatz} could prove very useful because of its similarity with the asymptotically-flat case. This could be performed both in three and four spacetime dimensions. In particular, it would be interesting to reproduce the analysis of \cite{Cotler:2024cia} in asymptotically-FLRW$_3$ spacetimes, since now the flux-balance law \eqref{mass scalar no Psi0 flux-balance} features a genuine soft term. The different orders at which radiation and Coulombic data enter the radial expansion for generic values of $k$ might appear as an obstacle towards the reconstruction of an infrared triangle. On the other hand, this characteristic already appears in asymptotically-flat gravity in more than four dimensions, where it did not hinder the identification of key features of the infrared triangle~\cite{Kapec:2015vwa,Pate:2017fgt}.
%%%%%%%%%%%%%%%%%%%%%%%%
\item\textbf{Wald--Zoupas prescription.}
One of the interesting aspects of the present model is that it enables to study, in a simplified setup, the subtleties appearing in the context of the generalized BMS group with arbitrary superrotations. This is what we have presented in subsection~\ref{sec:scalar_superrotations}. Although we could define charges that are conserved in the radiative vacuum defined by the vanishing of the covariant news, their bracket leads to a field-dependent cocycle. It would be very interesting to explore further the improved Wald--Zoupas prescription presented in \cite{Odak:2022ndm,Rignon-Bret:2024gcx,Rignon-Bret:2024wlu}, and to study the residual ambiguities in the definition of the charges~\eqref{scalar WZ charge with Y}. The question is whether further covariance criteria can be imposed in order to define charges whose bracket leads to a field-independent cocycle. In fact, this question already arises and is left unanswered in the case of three-dimensional asymptotically-flat spacetimes coupled to a Maxwell field \cite{Bosma:2023sxn}.
%%%%%%%%%%%%%%%%%%%%%%%%
\item\textbf{Cosmological history and epochs.}
Since observational evidence indicates that the present-day Universe is more accurately described by an FLRW geometry undergoing accelerated expansion \cite{SupernovaSearchTeam:1998fmf}, it is also important to study the asymptotic structure of such spacetimes. This is more complicated than in the decelerated case because of the different global structure, but techniques developed for dS asymptotics and for null horizons could play an important role. More generally, it would be interesting to understand the asymptotic structure of spacetimes which undergo a more complicated dynamics and go through several phases of energy domination, as in our cosmic history. Once again, this could be done in four spacetime dimensions, but non-trivial three-dimensional models certainly exist and may prove very informative as well.
%%%%%%%%%%%%%%%%%%%%%%%%
\end{itemize}

%%%%%%%%%%%%%%%%%%%%%%%%%%%%%%%%%%%%%%%%%%%%%%%%%%%%%%%%%%%%%%%%%%%

\section*{Acknowledgments}

We dedicate this work to the memory of Robert G. Leigh. We would like to thank Chrysoula Markou, Simone Speziale, Matthieu Vilatte and C\'eline Zwikel for useful discussions. The work of AC and NM was supported by the Fonds de la Recherche Scientifique -- FNRS under Grants nr.\ T.0047.24 and FC.60445. The work of AD was partially supported by the Belgian American Educational Foundation at the University of Illinois Urbana-Champaign, and by the Complexys Institute at the University of Mons. The work of MG is supported by the ANR PRC project “GeneraSyOn” (ANR-25-CE57-6085).

\appendix

%%%%%%%%%%%%%%%%%%%%%%%%%%%%
% New appendix style
\section*{Appendices}
\addcontentsline{toc}{section}{Appendices}
\renewcommand{\thesubsection}{\Alph{subsection}}
\setcounter{subsection}{0}
\counterwithin*{equation}{subsection}
\renewcommand{\theequation}{\Alph{subsection}.\arabic{equation}}
%%%%%%%%%%%%%%%%%%%%%%%%%%%%

\subsection{Three-dimensional asymptotically-flat spacetimes}
\label{app:vacuum flat}

In this appendix we briefly recall the structure of the solution space and of the residual symmetries for three-dimensional asymptotically-flat spacetimes \cite{Barnich:2006av,Barnich:2010eb,Barnich:2012aw}, both without and with matter (described by a scalar field). This enables to compare the boundary conditions and the solution space with the results for asymptotically-FLRW$_3$ spacetimes sourced by a stress-energy tensor. The ansatz for the line element in Bondi gauge is the same as \eqref{metric ansatz} with $k=0$, i.e.
\be\label{bondi ansatz}
\de s^2=\f{V}{r}e^{2\beta}\de u^2-2e^{2\beta}\de u\,\de r+r^2(\de\phi-U\de u)^2.
\ee

\paragraph{Absence of matter.}

The vacuum Einstein equations $E_{\mu\nu}\coloneqq G_{\mu\nu}=0$ can be organized into hypersurface, evolution, and trivial equations. First, $E_{rr}=0$ implies
\be
\beta(u,r,\phi)=\beta_0(u,\phi),
\ee
where $\beta_0$ is a radial integration constant. Next, $E_{r\phi}=0$ yields
\be
U(u,r,\phi)=U_0(u,\phi)+\f{2}{r}e^{2\beta_0}\beta_0'+\f{P(u,\phi)}{r^2},
\ee
where $U_0$ and $P$ are additional radial integration constants.\footnote{In parts of the literature, one instead uses the convention $P \to -P$ for the angular momentum aspect (see e.g.~\cite{Barnich:2010eb,Geiller:2021vpg}), so signs may differ accordingly in subsequent expressions.} Finally, $E_{ru}=0$, which is equivalent to $R_{\phi\phi}=0$, implies
\be
V(u,r,\phi)=-2r^2U_0'+2rM(u,\phi)-4P\beta_0'-\f{P^2}{r}e^{-2\beta_0},
\ee
where $M$ is another radial integration constant. One then finds that $E_{\phi\phi}=0$ is automatically satisfied as a consequence of the Bianchi identities.

The integration constants $M$ and $P$ play the role of mass and angular momentum aspects respectively. It is common to consider the boundary conditions
\be\label{flat vacuum BCs}
g_{uu}=\O(1),
\q\q
g_{ur}=-1+\O(r^{-1}),
\q\q
g_{u\phi}=\O(1).
\ee
The first of these boundary conditions follows from the absence of both a cosmological constant and a conformal Weyl factor in $g_{\phi\phi}$. Indeed, introducing a time-dependent Weyl factor $\varphi(u,\phi)$ through $g_{\phi\phi}=r^2 e^{2\varphi}$ would lead to $g_{uu}=\O(r)$, while a non-vanishing cosmological constant $\Lambda\neq 0$ would instead imply $g_{uu}=\O(r^2)$. The remaining two boundary conditions require setting $\beta_0=0=U_0$. These additional integration constants, beyond the mass and angular momentum aspects, act as boundary sources that obstruct the conservation of the asymptotic charges. Their role, together with that of the conformal Weyl factor discarded here, has been discussed in detail in~\cite{Ruzziconi:2020wrb,Geiller:2021vpg,Campoleoni:2022wmf,Ciambelli:2024vhy,Delfante:2025lxn,DArcy:2026hgl}. With $\beta_0=0=U_0$, the remaining Einstein equations $E_{uu}=0$ and $E_{u\phi}=0$ reduce respectively to the evolution equations
\be\label{flat vacuum M dot P dot}
\dot{M}=0,
\q\q
\dot{P}=-M'.
\ee
The asymptotic Killing vectors on $\I^+$ in the vacuum asymptotically-flat case take the form
\be\label{flat AKVs}
\xi_{(T,Y)}=f\partial_u+Y\partial_\phi,
\q\q
f=T+uY',
\ee
which corresponds to~\eqref{residual diffeo on scri} with $k=0$. The action of these vector fields on the mass and angular momentum is
\bsub\label{vacuum flat transformation laws}
\be
\delta_\xi M&=\big(f\partial_u+Y\partial_\phi+2Y'\big)M-Y''',\\
\delta_\xi P&=\big(f\partial_u+Y\partial_\phi+2Y'\big)P-2f'M+f'''.
\ee
\esub
This can be contrasted with the results obtained for asymptotically-FLRW spacetimes with a scalar field and no superrotations in \eqref{delta_scalar}, and with superrotations in \eqref{scalar transformations with Y}.

\paragraph{Scalar matter field.}

Let us now consider a non-trivial matter sector described by a scalar field~\cite{Ashtekar:1996cd,Cotler:2024cia}. The radial expansion for a radiative scalar field in Bondi coordinates is
\be\label{scalar expansion}
\Phi(u,r,\phi)=\sum_{n=0}^\infty\f{\Phi_n(u,\phi)}{r^{(2n+1)/2}},
\ee
where $\Phi_0(u,\phi)\eqqcolon C(u,\phi)$ is the unconstrained data which represents the scalar shear. This expansion therefore features half-integer powers of $r$. Using once again the line element \eqref{bondi ansatz}, and setting the radial integration constants $\beta_0=0=U_0$, the field equations are solved by
\bsub
\be
\beta&=-\f{1}{16r}C^2-\f{3}{16r^2}C\Phi_1+\O(r^{-3}),\\
U&=\f{P}{r^2}-\f{1}{24r^3}\big(2PC^2+3C'\Phi_1-5C\Phi_1'\big)+\O(r^{-4}),\\
V&=2rM+\f{1}{4}\big(CC''-(C')^2\big)+\O(r^{-1}).
\ee
\esub
Here the expansion of $\beta$ is determined by $E_{rr}=0$, that of $U$ by $E_{r\phi}=0$, and that of $V$ by $E_{\phi\phi}=0$. We have also already inserted the evolution equation for the subleading terms in \eqref{scalar expansion}, which arise from the Klein--Gordon equation. We have
\be\label{subleading scalar EOMs}
\Box\Phi=\O(r^{-(2n+5)/2})\q\Rightarrow\q\dot{\Phi}_{n\geq1}=(\,\dots),
\ee
and in particular
\bsub
\be
\dot{\Phi}_1&=\f{1}{4}\big(MC-2C''\big),\\
\dot{\Phi}_2&=\f{1}{64}\Big(72M\Phi_1-8P'C-32PC'+C(C')^2+5C^2C''-16\Phi_1''\Big).
\ee
\esub
Finally, the mass and the angular momentum satisfy
\be
\dot{M}=-\f{1}{2}N^2,
\q\q
\dot{P}=-M'+\f{1}{8}\big(3NC'-CN'\big),
\ee
which generalizes \eqref{flat vacuum M dot P dot}.

The asymptotic Killing vectors are given as in the vacuum case by \eqref{flat AKVs}, or equivalently by~\eqref{residual diffeo on scri} with $k=0$. They lead to the transformation laws
\bsub\label{scalar flat transformation laws}
\be
\delta_\xi M&=\big(f\partial_u+Y\partial_\phi+2Y'\big)M-Y''',\\
\delta_\xi P&=\big(f\partial_u+Y\partial_\phi+2Y'\big)P-2f'M+f'''-\f{1}{8}f'CN+\f{1}{16}Y''C^2,\\
\delta_\xi C&=\left(f\partial_u+Y\partial_\phi+\f{1}{2}Y'\right)C,\\
\delta_\xi N&=\left(f\partial_u+Y\partial_\phi+\f{3}{2}Y'\right)N,
\ee
\esub
where $N\coloneqq\dot{C}$. At the difference with the case of asymptotically-FLRW spacetimes, one can see that the fall-off conditions \eqref{scalar expansion} do not require an overleading term with respect to the shear in order to allow for non-trivial superrotations. Moreover, even with superrotations the news $N$ transforms homogeneously under the residual symmetries. Finally, one can see that $M$ and $P$ have the same conformal weight.

\subsection[Weyl--\texorpdfstring{$\mathfrak{bms}_3^k$ transformations}{bmsk3}]{Weyl--\texorpdfstring{$\boldsymbol{\mathfrak{bms}_3^k}$ transformations}{bmsk3}}
\label{app:Weyl}

In this appendix we extend the discussion of section \ref{sec:asymptotically-FLRW} to include Weyl transformations, and illustrate their intriguing role in mapping the Weyl--$\mathfrak{bms}_3^k$ algebra to Weyl--$\mathfrak{bms}_3$. For this, we generalize the metric ansatz \eqref{metric ansatz} to include a time-independent conformal factor $\varphi(\phi)$ in the angular part of the boundary metric. We therefore consider
\be\label{Weyl ansatz}
\de s^2=a^2\left(\f{V}{r}e^{2\beta}\de u^2-2e^{2\beta}\de u\,\de r+r^2e^{2\varphi}(\de\phi-U\de u)^2\right),
\q\q
a=(u+r)^k.
\ee
On top of this gauge fixing we then consider the same fall-off conditions as in \eqref{fall-offs metric}.

In order to find the residual diffeomorphisms, we have to take into account the Weyl conformal factor and replace the conditions \eqref{preserving the gauge} by
\be\label{preserving BS Weyl gauge}
\pounds_\xi g_{rr}=0,
\q\q
\pounds_\xi g_{r\phi}=0,
\q\q
\partial_r\left(\f{1}{a^2r^2}\pounds_\xi g_{\phi\phi}\right)=0.
\ee
This leads to
\bsub
\be
\xi^u&=f(u,\phi),\\
\xi^\phi&=y(u,\phi)-f'\int_r^\infty\f{e^{2\beta}}{\tilde{r}^2}\de\tilde{r},\\
\xi^r&=\f{r}{1+r\partial_r\ln a}\Big(W(\phi)+Uf'-f\partial_r\ln a-(\xi^\phi)'-\varphi'\xi^\phi\Big),
\ee
\esub
where $W(\phi)$ is a new free function entering the radial part of the vector field, and which appears because the last condition in \eqref{preserving BS Weyl gauge} is now differential and not just algebraic. It is constrained to depend only on $\phi$ since we have chosen the conformal factor to be $\varphi=\varphi(\phi)$ as well. With this vector field, preserving the fall-offs \eqref{fall-offs metric} first leads to
\be
\pounds_\xi g_{u\phi}=a^2\O(r)\q\Rightarrow\q y(u,\phi)=Y(\phi).
\ee
With this constraint we then find that the first fall-off condition is automatically preserved, i.e. $\pounds_\xi g_{uu}=a^2\O(1)$. Finally, preserving the second fall-off condition leads to
\be\label{f after constraint}
\pounds_\xi g_{ur}=a^2\O(r^{-1})\q\Rightarrow\q f(u,\phi)=T(\phi)+u\f{1+2k}{1+k}\big(Y'+Y\varphi'-W\big).
\ee
The upshot is that now the residual diffeomorphism is parametrized by three arbitrary functions on the circle, namely $T(\phi)$, $Y(\phi)$, and $W(\phi)$. The latter parametrizes the Weyl rescalings of the conformal factor, as can be seen from the transformation law $\delta_\xi\varphi=W$. In order to remove the field-dependence appearing in $\xi^u$ via the presence of $\varphi$ in \eqref{f after constraint}, one can perform a change of slicing by introducing
\be
Z(\phi)\coloneqq Y'+Y\varphi'-W.
\ee
At the end of the day, the residual vector field on the boundary takes the form
\be
\xi_{(T,Y,Z)}=\left(T+u\f{1+2k}{1+k}Z\right)\partial_u+Y\partial_\phi,
\ee
which should be contrasted with \eqref{residual diffeo on scri}. The relationship between these two results is that when setting $\varphi=0$ one should also impose $Z=Y'$.

We can now compute the Lie bracket of the above Weyl--BMS$_3^k$ vector fields, to find the algebraic structure
\be
\big[\xi_{(T_1,Y_1,Z_1)},\xi_{(T_2,Y_2,Z_2)}\big]=\xi_{(T_{12},Y_{12},Z_{12})},
\ee
with
\be
T_{12}\coloneqq Y_1T_2'+\f{1+2k}{1+k}T_1Z_2-(1\leftrightarrow 2),
\quad\
Y_{12}\coloneqq Y_1Y_2'-(1\leftrightarrow 2),
\quad\
Z_{12}\coloneqq Y_1Z_2'-(1\leftrightarrow 2).
\ee
Interestingly, this shows that upon redefining
\be
Z\to\f{1+k}{1+2k}Z
\ee
we obtain an isomorphism between Weyl--$\mathfrak{bms}_3^k$ and Weyl--$\mathfrak{bms}_3$. In other words, the presence of the Weyl freedom enables to reabsorb the dependence on $k$ which appears in the algebra. This is particularly intriguing since, when eliminating the Weyl component, $\mathfrak{bms}_3^k$ is not isomorphic to~$\mathfrak{bms}_3$. A similar observation was also made in the four-dimensional case in \cite{Enriquez-Rojo:2022onp}.

\subsection[Absence of translation subalgebra in \texorpdfstring{$\mathfrak{bms}_3^k$}{bmsk3}]{Absence of translation subalgebra in \texorpdfstring{$\mathfrak{bms}_3^k$}{bmsk3}}
\label{app:subalgebra}

In this appendix we show that when $Y(\phi)$ contains more than the constant mode $Y_0$ corresponding to $\mathfrak{so}(2)$ spatial rotations, then there is no translation subalgebra in $\mathfrak{bms}_3^k$. For this, let us first note that supertranslations form a Lie ideal of the algebra $\mathfrak{bms}_3^k$. Indeed, from the bracket~\eqref{Lie bracket} we obtain
\be
\big[\xi_T,\xi_Y\big]=\left(\f{1+2k}{1+k}TY'-YT'\right)\partial_u,
\ee
which is again a supertranslation, while two supertranslations commute. Since the generators of supertranslations and superrotations are functions on the circle $S^1$, they can be expanded in Fourier modes as
\be
T(\phi)=\sum_{n\in\mathbb{Z}}T_ne^{in\phi},
\q\q
Y(\phi)=\sum_{m\in\mathbb{Z}}Y_me^{im\phi},
\ee
in terms of which we have
\be\label{algebra in modes}
\big[\xi_{T_n},\xi_{Y_m}\big]=i\left(m\f{1+2k}{1+k}-n\right)T_nY_me^{i(n+m)\phi}\,\partial_u. 
\ee
Time translations correspond to $T_0$, while spatial translations correspond to $T_{\pm1}$. Similarly, spatial rotations correspond to $Y_0$, while Lorentz boosts are generated by $Y_{\pm1}$.

Let us now reduce $\mathfrak{bms}_3^k$ to the case where it contains only the constant rotations $Y(\phi)=Y_0$ and has trivial superrotations. This is indeed what we found in section \ref{subsec: scalar_stranslations} when studying the first choice of fall-off conditions for the scalar field. In this case we get from \eqref{algebra in modes} that
\be
\big[\xi_{T_n},\xi_{Y_0}\big]=-inT_nY_0e^{in\phi}\,\partial_u,
\ee
meaning that supertranslations do not mix and also that time translations are preserved. Note that this is a case which does not have a four-dimensional counterpart in the literature, since the analysis of \cite{Bonga:2020fhx} has revealed that $\mathfrak{bms}_4^k=\mathfrak{so}(1,3)\ltimes T_k(S^2)$, and there are no boundary conditions in this case which reduce the $\mathfrak{so}(1,3)$ part further
. It was shown in appendix A of~\cite{Bonga:2020fhx} that $\mathfrak{bms}_4^k$ has no translation subalgebra. Similarly, here we can already show that there is no translation subalgebra by considering an enlargement of $\mathfrak{so}(2)$ to $\mathfrak{so}(1,2)$. This will be sufficient in order to show that there is no translation subalgebra in $\mathfrak{bms}_3^k$ with non-trivial superrotations.

In addition to $Y_0$, let us therefore consider the Lorentz boosts and the algebra $\mathfrak{so}(1,2)$ generated by $\big(\xi_{Y_{-1}},\xi_{Y_0},\xi_{Y_{+1}}\big)$. We then find that it acts on time translations as
\be
\big[\xi_{T_0},\xi_{Y_{\pm 1}}\big]=\pm i\f{1+2k}{1+k}T_0Y_{\pm1}e^{\pm i\phi}\,\partial_u,
\q\q
\big[\xi_{T_0},\xi_{Y_0}\big]=0,
\ee
showing in particular that the Lorentz boosts map time translations to spatial translations. However, when acting with the boosts on the spatial translations we obtain
\bsub
\be
\big[\xi_{T_{+1}},\xi_{Y_{\pm1}}\big]&=\pm i\left(\f{1+2k}{1+k}\mp1\right)T_{+1}Y_{\pm1}e^{i(+1\pm1)\phi}\,\partial_u,\\
\big[\xi_{T_{-1}},\xi_{Y_{\pm1}}\big]&=\pm i\left(\f{1+2k}{1+k}\pm1\right)T_{-1}Y_{\pm1}e^{i(-1\pm1)\phi}\,\partial_u.
\ee
\esub
In the limit $k\to0$, which at the level of the algebra corresponds to the limit towards asymptotically-flat spacetimes since it leads to $\mathfrak{bms}_3$, there is a cancellation of terms and we recover as expected that the translations are preserved. This is in line with the fact that translations form a preferred subalgebra of all of $\mathfrak{bms}_3$ (although here we have looked at $\mathfrak{so}(1,2)$ and not all of $\mathrm{Vect}(S^1)$). When $k\neq0$, i.e. when the algebra is $\mathfrak{bms}_3^k$, we see that $k$ leads to the appearance of supertranslation modes (i.e. modes of the form $T_{\pm1}Y_{\pm1}$) on the right-hand side of the bracket. This prevents the translations $\xi_{T_{\pm1}}$ from forming a closed subalgebra of $\mathfrak{so}(1,2)\ltimes T_k(S^1)$, and thus of $\mathfrak{bms}_3^k=\mathrm{Vect}(S^1)\ltimes T_k(S^1)$.

\subsection{Einstein equations}
\label{app: Einstein eqs}

In this appendix we collect some lengthy explicit expressions for certain components of the Ricci tensor and of the Einstein equations $E_{\mu\nu}\coloneqq G_{\mu\nu}-T_{\mu\nu}=0$. Using a prime $'$ to denote $\partial_\phi$, and introducing
\be
\chi\coloneqq\f{r}{a^2}\partial_r\left(\f{a^2}{r}V\right)+\f{1}{2}\partial_rV-\f{1}{2}\partial_r\left(\f{r}{a^2}\partial_r(a^2V)\right),
\ee
we have
\bsub
\be
%%%%%%%%%%%%%%%%%%%%
R_{\phi\phi}
&=e^{-2\beta}\f{r}{a}\left(\partial_r\left(\partial_r(ar)\f{V}{r}\right)+\f{1}{ar}\partial_r\big(a^2r^2U'\big)+2\big(\partial_ua+r\partial_u\partial_ra\big)\right)\cr
&\pe-2\big(\beta'\big)^2-2\beta''-\f{1}{2}r^4e^{-4\beta}(\partial_rU)^2,\label{Rphiphi}\\
%%%%%%%%%%%%%%%%%%%%
E_{\phi\phi}
&=\f{2r^2}{a^2}e^{-2 \beta}\big(\partial_ra\partial_ua-a\partial_u\partial_ra\big)+\big(\beta'\big)^2-\f{3}{4}e^{-4\beta}r^4\big(\partial_rU\big)^2+e^{-2\beta}\chi\nn\\
&\pe-r^2e^{-2\beta}\left(\partial_r\left(\f{V}{r}\partial_r\beta\right)+\big(\beta'\partial_rU+2U\partial_r\beta'\big)+2\partial_u\partial_r\beta\right)-T_{\phi\phi},\label{Ephiphi}\\
%%%%%%%%%%%%%%%%%%%%
E_{u\phi}
&=-U\left\{\big(\beta'\big)^2+\beta''+e^{-2\beta}\left[\chi-2r^2\partial_r\big(\beta'U\big)+V\partial_r\beta-2r^2\partial_u\partial_r\beta+\f{r^2}{2}\partial_r U'-2r^2\partial_u\left(\f{1}{a}\partial_ra\right)\right]\right\}\nn\\
&\pe-\f{r}{2a}\partial_u\big(are^{-2 \beta}\partial_rU\big)-a\partial_u\left(\f{\beta'}{a}\right)-\f{\partial_r(aV')}{2ar}+\f{3}{4}e^{-4\beta}r^4 U(\partial_r U)^2\nn\\
&\pe+e^{-2\beta}\left(r\partial_r\big(UV\partial_r\beta\big)-r^2U'\partial_rU-\f{V}{2ar^2}\partial_r\big(ar^3\partial_rU\big)\right)+\f{V'}{r^2}\big(1-r\partial_r \beta\big)-T_{u\phi},\label{Euphi}\\
%%%%%%%%%%%%%%%%%%%%
E_{uu}&=-\f{1+r\partial_r\ln a}{2r^2}\partial_uV+(\,\dots)-T_{uu}.\label{Euu}
%%%%%%%%%%%%%%%%%%%%
\ee
\esub
These equations are valid for the line element \eqref{metric ansatz} with $k$ arbitrary. We do not give the explicit expression for $E_{uu}$ since it is extremely lengthy. Instead, we have only written the term involving $\partial_uV$ and which is responsible for the appearance of the mass loss formula at order $(ar)^{-1}$.

\subsection{Bare charges with superrotations}
\label{app:bare charges}

In this appendix we collect the expressions for the charges obtained in Einstein-scalar and Einstein--Maxwell theory when $\psi\neq0$ and with non-trivial superrotations.

\paragraph{Coupling to a scalar field.}

With an arbitrary choice of split between integrable and non-integrable terms, the $(ur)$ component of the Iyer--Wald charge \eqref{IW general Einstein scalar} is given by
\be\label{Iyer-Wald sc}
\oint\slashed{\delta}k^{ur}_\xi=r\slashed{\delta}Q_\xi^r+\delta Q_\xi+\Xi_\xi[\delta]+\O(r^{-1}),
\ee
where
\bsub\label{bare split IW scalar Y}
\be
%%%%%%%%%%%%%%%%%%%%%%%%%%%%%%%%%%%%%%%%%
\begin{split}
\slashed{\delta}Q_\xi^r
&=\f{1}{2}\oint f\Big(\delta\big((\psi')^2+\psi''\big)+\delta\psi\big(1+2\psi''+N\big)\Big)\\
&\pe\phantom{\f{1}{2}\oint}-\f{1}{2}Y\Big(\delta\big(2C\psi'+2C'+u\psi'\big)-\big(C\delta\psi\big)'\Big),
\end{split}\\
%%%%%%%%%%%%%%%%%%%%%%%%%%%%%%%%%%%%%%%%%
\begin{split}
Q_\xi
&=\f{1}{2}\oint f\left(4M+\f{1}{2}C\big(2-(\psi')^2-\psi''\big)+\f{u}{4}\big((\psi')^2-2\psi\big)\right)\cr
&\pe\phantom{\f{1}{2}\oint}-\f{1}{8}Y\Big(24P+(6u-10C)C'+\big(6u^2-6uC-5C^2\big)\psi'\Big),
\end{split}\\
%%%%%%%%%%%%%%%%%%%%%%%%%%%%%%%%%%%%%%%%%
\begin{split}
\Xi_\xi[\delta]
&=\f{1}{2}\oint f\left(N\delta C+\f{1}{2}\big(\psi''-3(\psi')^2\big)\delta C-2\psi'\delta C'+\f{1}{2}\delta\psi'\psi'C\right)\cr
&\pe\phantom{\f{1}{2}\oint}+f\delta\psi\left(4M+\f{u}{2}N+C''-C\psi''-\f{1}{2}(u+C)(\psi')^2\right)+\f{1}{4}Yu\big(C\delta\psi\big)'.
\end{split}
%%%%%%%%%%%%%%%%%%%%%%%%%%%%%%%%%%%%%%%%%
\ee
\esub
One can see in particular that the charge has a divergent contribution. For this particular split of the charges into integrable and non-integrable parts, the cocycle produced by the Barnich--Troessaert bracket \eqref{BT bracket} is
\be\label{cocycle IW scalar Y}
\begin{split}
\K_{\xi_1,\xi_2}
&=\f{1}{16}\oint4C\Big(\psi'f_1Y_2''+2f_1Y_2'''-4f_1Y_2'+4f_1'Y_2''\Big)-12uf_1\big(Y_2'+Y_2'''\big)\cr
&\pe\phantom{\f{1}{16}\oint}+2\psi\left(uY_1'f_2''-4f_1f_2^{(4)}+uf_1\big(5Y_2'+Y_2'''\big)\right)-(\psi')^2f_1\big(3uY_2'+2f_2''\big)\cr
&\pe\phantom{\f{1}{16}\oint}+4\big(4\dot{C}\psi'-3(\psi')^3\big)f_1f_2'-(1\leftrightarrow2),
\end{split}
\ee
and is manifestly field-dependent.

\paragraph{Coupling to a Maxwell field.}

With an arbitrary choice of split between integrable and non-integrable terms, the $(ur)$ component of the Iyer--Wald charge \eqref{Maxwell IW charge} is given by
\be\label{bare split IW Maxwell 1}
\oint\slashed{\delta}k^{ur}_{(\xi,\epsilon)}=\delta Q_{(\xi,\epsilon)}+\Xi_\xi[\delta]+\O(r^{-1}),
\ee
where
\bsub\label{bare split IW Maxwell 2}
\be
%%%%%%%%%%%%%%%%%%%%%%%%%%%%%%%%%%%%%%%%%
\begin{split}
Q_{(\xi,\epsilon)}
&=\oint\alpha E+\f{1}{8}f\Big(16M+5C+uE^2-2CE^2+2uE'-2EC'\Big)\\
&\qquad-\f{1}{16}Y\Big(24P-\big(16A_\phi^0-8u^2+6uC+5C^2\big)E-16\big(A_\phi^0\big)'+2C'(3u-5C)\Big),
\end{split}\\
%%%%%%%%%%%%%%%%%%%%%%%%%%%%%%%%%%%%%%%%%
\Xi_\xi[\delta]&=\f{1}{2}\oint f\left(2A_u^0\delta E-\f{3}{2}E\big(E\delta C+\delta C'\big)+\dot{C}\delta C\right).
%%%%%%%%%%%%%%%%%%%%%%%%%%%%%%%%%%%%%%%%%
\ee
\esub
At the difference with the scalar case, the charge is now manifestly finite. For this particular split of the charges into integrable and non-integrable parts, the cocycle produced by the Barnich--Troessaert bracket \eqref{BT bracket} is
\be\label{Coc Einstein-Maxwell}
\begin{split}
\K_{(\xi_1,\alpha_1),(\xi_2,\alpha_2)}
&=-\f{1}{16}\oint f_1\Big(16(u+C)+3uE^2\Big)Y_2'\cr
&\pe\phantom{-\f{1}{16}\oint}+2f_1\Big(8uY_2'''-3CEY_2''-3E^2f_2''+8E\dot{\alpha}_2+f_2'E\big(2+6E^2-6\dot{C}\big)\Big)\cr
&\pe\phantom{-\f{1}{16}\oint}+2f_1\Big(6CY_2''+E\big(uY_2'-6f_2''\big)\Big)'+16\alpha_1Y_2''-(1\leftrightarrow2),
\end{split}
\ee
and is also manifestly field-dependent as in the case of scalar matter.

\subsection[Vacuum structure for \texorpdfstring{$k=1$}{k=1}]{Vacuum structure for \texorpdfstring{$\boldsymbol{k=1}$}{k=1}}
\label{Appendix vacuum}

To identify the structure of the asymptotic symmetry orbit containing the FLRW background with~$k =1$, we start from a parametrization of the background in a given coordinate system and perform a general finite diffeomorphism preserving the gauge-fixing conditions \eqref{metric ansatz} and the fall-offs \eqref{fall-offs beta V u}.\footnote{Starting from Bondi coordinates would generate diffeomorphisms that spoil the possibility of matching our vacua with our radial expansion, and thus with our solution space. We choose conformal coordinates detailed in section \ref{sec:exactly-FLRW} for technical convenience.} As mentioned in subsection~\ref{sec:vacuum}, we start from the background solution in conformal coordinates,
\bsub\label{gstart app}
\be
\de s^2&=\tilde{g}_{\mu\nu}\de\tilde{x}^\mu\de\tilde{x}^\nu=a(\tau)^2\big(\!-\!\de\tau^2+\de\rho^2+\rho^2\de\theta^2\big),\qquad a(\tau)=\tau,\\
\Phi(\tau)&=2\ln(\tau),
\ee
\esub
and we act on it with an arbitrary finite diffeomorphism~$\tilde{x}^\mu=(\tau,\rho,\theta)\mapsto x^\mu=(u,r,\phi)$. We then impose that the transformed metric
\be
g_{\mu \nu}(x)=\f{\partial\tilde{x}^\alpha}{\partial x^\mu}\f{\partial\tilde{x}^\beta}{\partial x^\nu}\tilde{g}_{\alpha\beta}\big(\tilde{x}(x)\big)
\ee
satisfies the conditions \eqref{metric ansatz} and \eqref{fall-offs beta V u} with $k=1$. 

Motivated by the relation between conformal time and retarded conformal time, as well as by the structure of our solution space (which admits an expansion in integer powers of $r$), we assume that the conformal coordinates can be expanded as
\bsub\label{new coord}
\be
\tau(u, r, \phi)&=\tau_{-1}(u,\phi)r+\sum_{n=0}^\infty\f{\tau_n(u,\phi)}{r^n},\\
\rho(u, r, \phi)&=\rho_{-1}(u,\phi)r+\sum_{n=0}^\infty\f{\rho_n(u,\phi)}{r^n},\\
\theta(u, r, \phi)&=\sum_{n=0}^\infty\f{\theta_n(u,\phi)}{r^n}.
\ee
\esub
Imposing the gauge-fixing conditions \eqref{metric ansatz}, i.e. $g_{rr}=0$, $g_{r\phi}=0$, and $g_{\phi\phi}=a^2r^2=(r+u)^2r^2$, leads respectively to the equations
\bsub
\be
\left(\f{\partial\tau}{\partial r}\right)^2&=\left(\f{\partial\rho}{\partial r}\right)^2+\rho^2\left(\f{\partial\theta}{\partial r}\right)^2,\label{g'rr}\\
\left(\f{\partial\tau}{\partial r}\right)\left(\f{\partial\tau}{\partial\phi}\right)&=\left(\f{\partial\rho}{\partial r}\right)\left(\f{\partial\rho}{\partial\phi}\right)+\rho^2 \left(\f{\partial \theta}{\partial r}\right)\left(\f{\partial\theta}{\partial\phi}\right),\label{g'rphi}\\
(r+u)^2r^2&=\tau^2\left[-\left(\f{\partial\tau}{\partial\phi}\right)^2+\left(\f{\partial\rho}{\partial\phi}\right)^2+\rho^2\left(\f{\partial \theta}{\partial\phi}\right)^2\right],\label{g'phiphi}
\ee
\esub
which can be solved iteratively in order to determine the radial expansions of the coordinates $\tilde{x}^\mu=(\tau,\rho,\theta)$. At leading orders, together with the boundary conditions \eqref{fall-offs metric}, they yield
\bsub\label{sol finite diffeo}
\begin{alignat}{5}
&a^{-2}g_{rr}=\O(r^{-1})&\qquad\Rightarrow\qquad&\rho_{-1}(u,\phi)=\tau_{-1}(u,\phi),\\
&a^{-2}g_{uu}=\O(r)&\qquad\Rightarrow\qquad&\theta_0(u,\phi)=\theta_0(\phi),\\
&a^{-2}g_{\phi\phi}-r^2=\O(r)&\qquad\Rightarrow\qquad&\tau_{-1}(u,\phi)=\tau_{-1}(\phi)=\big(\theta_0'\big)^{-1/2},\label{orientation cond}\\
&a^{-2}g_{ur}+1=\O(r^{-1})&\qquad\Rightarrow\qquad&\tau_0(u,\phi)=\big(\theta_0'\big)^{3/2}u+\rho_0(u,\phi)+\vartheta_0(\phi),\\
&a^{-2}g_{r\phi}=\O(r^{-1})&\qquad\Rightarrow\qquad&\theta_1(u,\phi)=\big(\theta_0'\big)^{-1/2}\vartheta_0'-\frac{3}{2}u\theta_0'',\\
&a^{-2}g_{\phi\phi}-r^2=\O(1)&\qquad\Rightarrow\qquad&\rho_0(u,\phi)=\bar{\rho}_0(u,\phi),
\end{alignat}
\esub
where
\be\label{R0}
\begin{split}
\bar{\rho}_0(u,\phi)=\f{1}{8\big(\theta_0'\big)^3}
\Big(&4u\big(\theta_0'\big)^{5/2}-4u\big(\theta_0'\big)^{9/2}-3u\big(\theta_0'\big)^{1/2}\big(\theta_0''\big)^2+6u\big(\theta_0'\big)^{3/2}\theta_0'''\\
&+4\theta_0'\vartheta_0''-4\theta_0''\vartheta_0'-4\vartheta_0\big(\theta_0'\big)^3\Big).
\end{split}
\ee
Up to these orders, all quantities only depend on $\theta_0=\theta_0(\phi)$ and $\vartheta_0=\vartheta_0(\phi)$, which are free angular functions. In \eqref{orientation cond}, we imposed that $\theta_0'$ is a positive function, which is equivalent to requiring that the finite diffeomorphism preserves orientation. All subleading orders in the expansions of the coordinates $\tilde{x}^\mu=(\tau,\rho,\theta)$ are then fully determined by the gauge-fixing equations \eqref{g'rr}, \eqref{g'rphi}, and \eqref{g'phiphi}. They can be solved iteratively order by order according to the following scheme: the condition $a^{-2}g_{rr}=\O(r^{-(n+2)})$ fixes $\rho_n$ in terms of $\theta_0$, $\vartheta_0$, and $\tau_n$, while $a^{-2}g_{r\phi}=\O(r^{-n})$ and $a^{-2}g_{\phi\phi}-r^2=\O(r^{-n})$ determine $\theta_n$ and $\tau_n$, respectively, in terms of $\theta_0$ and $\vartheta_0$.

In conclusion, the two angular functions $\theta_0$ and $\vartheta_0$ completely determine the allowed finite diffeomorphisms. We therefore obtain a two-parameter family of vacua, parametrized by $\theta_0$ and $\vartheta_0$. It is instructive to compare this result with the residual symmetry algebra \eqref{residual diffeo on scri}, which is spanned by supertranslations and superrotations. To this end, we note that combining the transformation laws \eqref{residual diffeo on scri} of the metric components with \eqref{sol finite diffeo} gives
\bsub
\be
\delta_\xi\theta_0'&=Y\partial_\phi\theta_0'+\theta_0'Y',\\
\delta_\xi\vartheta_0&=Y\partial_\phi\vartheta_0+\big(\theta_0'\big)^{3/2}T.
\ee
\esub
This suggests to introduce the quantities $\T(\phi)$ and $\W(\phi)$ defined as
\bsub\label{Notation W}
\be
\theta_0'&\eqqcolon e^{\W},\\
\vartheta_0&\eqqcolon e^{3\W/2}\T,
\ee
\esub
and which then transform as
\bsub
\be
\delta_\xi\W&=Y\W'+Y',\\
\delta_\xi\T&=\left(Y\partial_\phi-\f{3}{2}Y'\right)\T+T,
\ee
\esub
where we used that $\theta_0'$ in \eqref{Notation W} is a strictly positive function, as we asked to preserve orientation. This is the same transformation as that of the generators of the residual symmetries, cf.~\eqref{eq. bms3like}.

In terms of this new parametrization, the action of the finite diffeomorphism \eqref{sol finite diffeo} on the background \eqref{gstart app} yields
\bsub
\be\label{gfin app}
\de s^2
&=g_{\mu\nu}\de x^\mu\de x^\nu\cr
&=a^2\Bigg[\left(-1+\frac{3}{2}\big(\W'\big)^2-\frac{3}{2}\W''+\O(r^{-1})\right)\de u^2+2\big(\!-\!1+\O(r^{-1})\big)\de u\,\de r\cr
&\pe\pe\pe+2\big(\!-\!\W'r+\O(1)\big)\de u\,\de\phi+r^2\de\phi^2\Bigg],\\
\Phi(x)&=2\ln(r+u)-\W(\phi)+\O(r^{-1}),
\ee
\esub
which eventually leads to \eqref{vacuum-valued bondi param}. 

\bibliography{Biblio.bib}
\bibliographystyle{Biblio}

\end{document}